\pdfoutput=1
\documentclass[11pt,a4paper]{article}
\usepackage{emnlp2022}
\usepackage{times}
\usepackage{latexsym}

\usepackage{multicol}
\usepackage{amsmath}
\usepackage{amsthm}
\usepackage{amssymb}
\usepackage{amsfonts}
\usepackage{latexsym}
\usepackage{rotating}
\usepackage{adjustbox}
\usepackage{relsize}
\usepackage{amsthm}
\usepackage{booktabs}
\usepackage{adjustbox}
\usepackage{booktabs}
\usepackage{url}
\usepackage{bbm}
\usepackage{scalerel}
\usepackage{pifont}
\usepackage{MnSymbol}

\usepackage{inconsolata}  

\usepackage{algorithm}
\usepackage[noend]{algpseudocode}
\newcommand{\alglinenumNew}[1]{\newcounter{ALG@line@#1}}
\newcommand{\alglinenumSave}[1]{\setcounter{ALG@line@#1}{\value{ALG@line}}}
\newcommand{\alglinenumRestore}[1]{\setcounter{ALG@line}{\value{ALG@line@#1}}}
\algdef{SE}[SUBALG]{Indent}{EndIndent}{}{\algorithmicend\ }%
\algtext*{Indent}
\algtext*{EndIndent}

\usepackage{subcaption}
\usepackage{graphicx}
\usepackage[font=small]{caption}
\usepackage{microtype}
\usepackage{mathtools} 

\usepackage{color}
\usepackage{xcolor}
\definecolor{darkgrey}{rgb}{0.2,0.2,0.2}
\definecolor{grey}{rgb}{0.9,0.9,0.9}
\definecolor{darkblue}{rgb}{0.0,0.0,0.5}
\definecolor{darkred}{rgb}{0.5,0.0,0.0}
\definecolor{darkorange}{rgb}{1.0,0.55,0.0}
\definecolor{darkgreen}{rgb}{0.0,0.6,0.0}
\definecolor{darkyellow}{rgb}{1.0,0.65,0.0}
\definecolor{darkorange}{rgb}{1.0,0.65,0.0}
\definecolor{darkergreen}{rgb}{0.0,0.4,0.0}
\definecolor{lightblue}{rgb}{0.8,0.8,1.0}
\definecolor{lightgreen}{rgb}{0.8,1.0,0.8}
\definecolor{lightred}{rgb}{1.0,0.8,0.8}
\definecolor{lightyellow}{rgb}{1.0,1.0,0.8}
\definecolor{lightorange}{rgb}{1.0,0.9,0.8}
\definecolor{lightgrey}{rgb}{0.96,0.97,0.98}
\definecolor{brilliantlavender}{rgb}{0.96, 0.73, 1.0}

\definecolor{ryanred}{rgb}{0.64, 0.0, 0.0}
\definecolor{ryanblue}{rgb}{0.13, 0.0, 0.58}
\definecolor{ryangreen}{rgb}{0.12, 0.59, 0.0}
\definecolor{ryanpurple}{rgb}{0.65, 0.0, 0.57}

\definecolor{mylavender}{HTML}{BD71E1}
\definecolor{darkpurple}{HTML}{531B93}

\usepackage{xcolor}
\usepackage{color}

\usepackage{comment}
\newcommand{\saveForCR}[1]{}

\usepackage[disable]{todonotes}
\makeatletter
\newcommand*\iftodonotes{\if@todonotes@disabled\expandafter\@secondoftwo\else\expandafter\@firstoftwo\fi} 
\makeatother

\newcommand{\note}[4][]{\todo[author=#2,color=#3,size=\scriptsize,fancyline,caption={},#1]{#4}} 
\newcommand{\ryan}[2][]{\note[#1]{Ryan}{violet!40}{#2}}

\newcommand{\anej}[2][]{\note[#1]{Anej}{darkgreen!40}{#2}}

\newcommand{\jason}[2][]{\note[#1]{Jason}{red!40}{#2}}
\newcommand{\Jason}[2][]{\jason[inline,#1]{#2}}
\newcommand{\benji}[2][]{\note[#1]{Benji}{orange!40}{#2}}

\newcommand{\response}[1]{\vspace{3pt}\hrule\vspace{3pt}\textbf{#1:}}
\newlength{\extramargin}
\usepackage{calc}

\usepackage{cleveref}

\crefname{section}{\S}{\S\S}
\Crefname{section}{\S}{\S\S}
\crefname{table}{Table}{Tables}
\crefname{figure}{Fig.}{Figs.}
\crefname{algorithm}{Alg.}{Algs.}
\crefname{equation}{Eq.}{Eqs.}
\crefname{line}{line}{lines}
\crefname{appendix}{App.}{Appendices}
\crefformat{section}{\S#2#1#3}  

\usepackage{tikz}
\usetikzlibrary{automata, fit, positioning, matrix, arrows.meta, decorations.pathmorphing, shapes.multipart}

\tikzset{
  heap/.style={
    every node/.style={circle,draw},
    level 1/.style={sibling distance=30mm},
    level 2/.style={sibling distance=10mm}
  }
}
\tikzset{
-{Stealth[length=2mm, width=2mm]}, 
node distance=2cm,
every state/.style={thick, fill=gray!10},
initial text=$ $,
}

\newcommand{\algorithmicfunc}[1]{\textbf{def} {#1}:}
\algdef{SE}[FUNC]{Func}{EndFunc}[1]{\algorithmicfunc{#1}}{}
\makeatletter
\ifthenelse{\equal{\ALG@noend}{t}}%
  {\algtext*{EndFunc}}
  {}%
\makeatother

\renewcommand{\algorithmicindent}{9pt}
\makeatletter
\newcommand{\algmargin}{\the\ALG@thistlm}
\makeatother
\algnewcommand{\Statepar}[1]{\State\parbox[t]{\dimexpr\linewidth-\algmargin}{\strut #1\strut}}
\algnewcommand{\LineComment}[1]{\State {\color{black!50!green}\smaller \(\triangleright\) \parbox[t]{\linewidth-\leftmargin-\widthof{\(\triangleright\) }}{\it #1}\smallskip}} 
\algnewcommand{\InlineComment}[1]{\hfill {\color{black!50!green}\(\triangleright\) {\tiny \it #1}}}
\algrenewcommand\algorithmicindent{1.0em}%

\newcommand{\defn}[1]{\textbf{#1}}

\DeclareMathOperator*{\mean}{mean}

\newcommand{\bigO}{\mathcal{O}}

\newcommand{\opluseq}{\mathrel{\oplus\!\!=}}

\newcommand{\monsteroplus}[2][lr]{\smashoperator[#1]{\bigoplus_{\substack{#2}}}}

\newcommand{\alphabet}{{ \Sigma}}
\newcommand{\states}{{ Q}}
\newcommand{\qchild}{{ q'}}

\newcommand{\transitions}{{ E}}
\newcommand{\failureexpansion}{{ \overline{\transitions}}}
\newcommand{\initf}{{ \lambda}}
\newcommand{\finalf}{{ \rho}}
\newcommand{\failuref}{{ \phi}}
\newcommand{\wfsa}{{ \mathcal{A}}}

\newcommand{\wfsatuple}{{ \langle \alphabet, \states, \transitions, \initf, \finalf \rangle}}
\newcommand{\failurewfsatuple}{{ \langle \alphabet, \states, \transitions, \initf, \finalf, \failuref \rangle}}
\newcommand{\updatecomplexity}{C_U}

\newcommand{\apath}{{ \boldsymbol \pi}}
\newcommand{\paths}{{ \Pi}}
\newcommand{\weight}{{ \mathrm{w}}}
\newcommand{\innerweight}{{ \weight_{\mathrm{I}}}}
\newcommand{\prevq}{{ \mathrm{p}}}
\newcommand{\nextq}{{ \mathrm{n}}}
\newcommand{\fallbackstate}{{ q^{\phi}}}
\newcommand{\ancstate}{q'}
\newcommand{\ancs}{\mathrm{ancs}}
\newcommand{\avgoutfrac}{{ s}} 

\newcommand{\failurepath}{{ \apath}}
\newcommand{\maxfailurepath}{{ \failurepath_{\mathrm{max}}}}
\newcommand{\maxfailuretree}{{ \tree_{\mathrm{max}}}}
\newcommand{\maxfailurepathsize}{{ |\maxfailurepath|}}
\newcommand{\maxfailuretreesize}{{ |\maxfailuretree|}}

\newcommand{\edge}[4]{{#1 \xrightarrow{#2 / #3} #4}}

\newcommand{\outsymbols}{{ \alphabet}}
\newcommand{\extendedoutsymbols}{{ \overline{\alphabet}}}

\newcommand{\qroot}{{ q_{\textit{root}}}}

\newcommand{\tree}{{ \mathcal{T}}}

\newcommand{\pathsum}{{ \mathbf{Z}}}
\newcommand{\backwardweight}{{ \beta}}

\newcommand{\aggregator}{{ \boldsymbol{\gamma}}}
\newcommand{\oldaggregator}{{ \aggregator_{\text{old}}}}
\newcommand{\updirection}{{ \uparrow}}
\newcommand{\downdirection}{{ \downarrow}}

\newcommand{\semiring}{{ \mathcal{W}}}

\newcommand{\semiringset}{{ \mathbb{K}}}
\newcommand{\semizero}{{\textbf{0}}}
\newcommand{\semione}{{\textbf{1}}}
\newcommand{\semiringtuple}{{\left(\semiringset, \oplus, \otimes, \semizero, \semione \right)}}
\newcommand{\additiontuple}{{\left(\semiringset, \oplus, \semizero \right)}}
\newcommand{\multiplicationtuple}{{\left(\semiringset, \otimes, \semione \right)}}

\newcommand{\ignore}[1]{}

\usepackage{relsize}
\newcommand{\myfunc}[1]{\textsf{\smaller\color{blue!50!black}#1}}

\colorlet{phitransitioncolor}{orange!80!black}
\colorlet{failureexpansioncolor}{purple!80!black}
\colorlet{localtermcolor}{cyan!55!black}
\colorlet{failuretermcolor}{failureexpansioncolor}
\newcommand{\localterm}[1]{{\color{localtermcolor}#1}}
\newcommand{\failureterm}[1]{{\color{failuretermcolor}#1}}

\newcommand{\backward}{\myfunc{Backward}}
\newcommand{\genericbackward}{\myfunc{GeneralBackward}}
\newcommand{\memoizationbackward}{{\myfunc{MemoizationBackward}}}
\newcommand{\ringbackward}{{\myfunc{RingBackward}}}
\newcommand{\failureexpansionprocedure}{\myfunc{FailureExpansion}}

\newcommand{\updatestep}{\myfunc{Visit}}
\newcommand{\updatesteps}{\myfunc{Visit\textsuperscript{+}}}
\newcommand{\restorestep}{\myfunc{Leave}}
\newcommand{\reversetopo}{\myfunc{ReverseTopological}}

\newcommand{\aggregatortype}{\myfunc{Aggregator}}
\newcommand{\findpath}{\myfunc{Path}}
\newcommand{\previousexpandedstate}{\myfunc{Prev}}

\usepackage{pgf}
\usepackage[utf8]{inputenc}
\usepackage{xspace}
\usepackage{enumitem}

\usepackage{array}
\newcolumntype{P}[1]{>{\centering\arraybackslash}p{#1}}

\newcommand{\defeq}[0]{\mathrel{\stackrel{\textnormal{\tiny def}}{=}}}

\newtheorem{defin}{Definition}
\newtheorem{theorem}{Theorem}[section]
\newtheorem{lemma}{Lemma}[section]

\newcommand{\ethz}{{ \mathrm{1}}}
\newcommand{\jhu}{{ \mathrm{2}}}

\title{Algorithms for Acyclic Weighted Finite-State Automata with Failure Arcs}
\author{
Anej Svete$^{\ethz}$~\;~Benjamin Dayan$^{\ethz}$ \\ \textbf{Tim Vieira$^{\jhu}$~\;~Ryan Cotterell$^{\ethz}$~\;~Jason Eisner$^{\jhu}$}\\
$^{\ethz}$ETH Z\"{u}rich
~\;~$^{\jhu}$Johns Hopkins University \\
  \texttt{\{\href{mailto:asvete@ethz.ch}{asvete}, \href{mailto:bdayan@ethz.ch}{bdayan}\}@ethz.ch} \\
  \texttt{\href{mailto:ryan.cotterell@inf.ethz.ch}{ryan.cotterell}@inf.ethz.ch} ~\;~
  \texttt{\{\href{mailto:timv@cs.jhu.edu}{timv}, \href{mailto:jason@cs.jhu.edu}{jason}\}@cs.jhu.edu} 
}

\date{}

\begin{document}

\setlength{\abovedisplayskip}{3pt}
\setlength{\belowdisplayskip}{3pt}

\maketitle

\begin{abstract}
Weighted finite-state automata (WSFAs) are commonly used in NLP. 
Failure transitions are a useful extension for compactly representing backoffs or interpolation in $n$-gram models and CRFs, which are special cases of WFSAs. 
The pathsum in ordinary acyclic WFSAs is efficiently computed by the backward algorithm in time $\bigO{}(|\transitions|)$, where $\transitions$ is the set of transitions.
However, this does not allow failure transitions, and preprocessing the WFSA to eliminate failure transitions could greatly increase $|\transitions|$.
We extend the backward algorithm to handle failure transitions directly.
Our approach is efficient when the average state has outgoing arcs for only a small fraction $\avgoutfrac \ll 1$ of the alphabet $\alphabet$.
We propose an algorithm for general acyclic WFSAs which runs in $\bigO{\left(|\transitions| + \avgoutfrac |\alphabet| |\states| \maxfailuretreesize \log{|\alphabet|}\right)}$, where $\states$ is the set of states and $\maxfailuretreesize$ is the size of the largest connected component of failure transitions.
When the failure transition topology satisfies a condition exemplified by CRFs,
the $\maxfailuretreesize$ factor can be dropped, and when the weight semiring is a ring, the $\log{|\alphabet|}$ factor can be dropped.  In the latter case (ring-weighted acyclic WFSAs), we also give an alternative algorithm with complexity $\displaystyle \bigO{\left(|\transitions| + |\alphabet| |\states| \min(1,s\maxfailurepathsize) \right)}$, where $\maxfailurepathsize$ is the size of the longest failure path.\looseness=-1

\vspace{0.5em}
\hspace{.5em}\includegraphics[width=1.25em,height=1.25em]{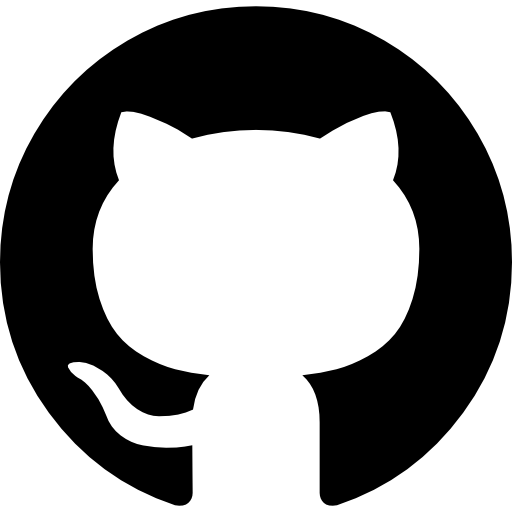}\hspace{.75em}\parbox{\dimexpr\linewidth-2\fboxsep-2\fboxrule}{\url{https://github.com/rycolab/failure-backward}}
\vspace{-.5em}
\end{abstract}

\section{Introduction}

Weighted finite-state automata (WFSAs) are a common formalism in NLP.
Many popular models are special cases, e.g., $n$-gram language models \citep{brown-etal-1992-class}, conditional random fields \cite[CRFs:][]{DBLP:conf/icml/LaffertyMP01}, maximum-entropy Markov models \cite{McCallum2000MaximumEM}, and semi-Markov models \citep{NIPS2004_eb06b9db}.\ryan{This is only true if the span length is bounded, I think. Should we have a footnote?}
In current practice, the weights in the WFSAs are often derived from a neural network, and neuralized WFSAs constitute the state of the art on a variety of common tasks in NLP \citep{rastogi-etal-2016-weighting, schwartz-etal-2018-bridging, lin-etal-2019-neural, jiang-etal-2021-neuralizing, 10.1162/tacl_a_00427, https://doi.org/10.48550/arxiv.2201.12431}.\ryan{I have a nostalgic urge to cite my first NLP paper from 2014.}
WFSAs are also increasingly being used for the design \cite{https://doi.org/10.48550/arxiv.1810.09536, schwartz-etal-2018-bridging} and analysis \cite{peng-etal-2018-rational, hewitt-etal-2020-rnns, 10.1162/tacl_a_00306, chiang-cholak-2022-overcoming} of neural architectures.\looseness=-1

\defn{Failure transitions} are a useful augmentation of standard WFSAs.
First introduced in the context of string matching \citep{aho-corasick-1975}, they can be used to represent backoff $n$-gram language models \citep{allauzen-mohri-roark:2003:ACL}, higher-order CRFs, and variable-order CRFs \citep[VoCRFs;][]{vieira-cotterell-eisner:2016:EMNLP2016} in a more compact way.
They represent ``default'' transitions out of states when no other transition is possible.
For example, in backoff $n$-gram language models, a weighted failure transition from a higher-order history to a lower-order history (e.g., from a 4-gram to a 3-gram) is used to back off before reading a word that was rarely observed with the higher-order history, so that it was not worth including a dedicated transition for that word.\looseness=-1
%

The \defn{pathsum} computes the total weight of all the paths in a WFSA graph, where the weights may fall in any semiring.\footnote{We formally define the problem in \cref{sec:preliminaries}.}
Examples include finding the highest-weighted path for Viterbi decoding, computing the posterior marginals (inference) in hidden Markov models, and computing the normalizing constant in CRFs.
The pathsum is particularly efficient to compute in \emph{acyclic} WFSAs with the \emph{backward algorithm}, whose runtime is $\bigO{\left(|\transitions|\right)}$.
However, the special semantics of failure transitions mean that the ordinary backward algorithm cannot be applied (nor can the forward algorithm).
Failure transitions must first be replaced by normal ones (\cref{alg:failure-expansion} below), resulting in the failure-expanded transition set $\failureexpansion$, which can contain up to  $|\states|^2|\alphabet|$ transitions. 
Replacing failure transitions, therefore, undoes the compaction afforded by them.
This is especially expensive ($|\failureexpansion|\gg|\transitions|$) for backoff language models, for example, where each of the many 4-gram states only has explicit transitions in $\transitions$ for symbols $a$ that were observed in training data to follow that 4-gram, but has transitions in $\failureexpansion$ for every $a \in \alphabet$.
For example, Penn Treebank tagging has $|\alphabet|=36$ and Czech morphological tagging has $|\alphabet|\! >\! 1000$ \cite{hajic-hladka-1998-tagging}.
While \citet{allauzen-mohri-roark:2003:ACL} present an $\bigO{}(n^2|\alphabet||\states|)$ method to preprocess a (possibly cyclic) $n$-gram language model WFSA with failure transitions such that the pathsum remains identical, their method only applies to the case of the tropical semiring.\jason{and only to n-gram models?  But \response{Anej}{Yes, only n-gram models} \response{jason}Doesn't it also assume that an path using explicit transitions (when available) will have a better weight than a path that backs off?\response{anej}{huh, I'm not sure... I don't remember anything about that from the paper explicitly, but it may have just been hidden in he fact they use the tropical semiring}} \looseness=-1 

In this paper, we study the problem of  efficiently computing the pathsum in WFSAs with failure transitions over general semirings. 
We specifically focus on \emph{acyclic} WFSAs,\footnote{\label{fn:cyclic}We have also worked out several novel algorithms for \emph{cyclic} WFSAs with failure transitions. However, the $9$-page format of EMNLP submissions meant that we had to save these for a future manuscript. Moreover, we view the split into acyclic and cyclic cases as natural, as our algorithms for cyclic automata are based on the more complex algorithm of \citet{lehmann-1977}.
While the acyclic algorithms cannot be run on a backoff language model (one of our examples), they can be used, for example, to compute the total language model probability of all paths in an acyclic lattice that may have wildcard arcs and/or failure arcs. Even though the language model is cyclic, its intersection with the lattice becomes acyclic.} introducing several algorithms, all based on the backward algorithm, that take advantage of the compact structure induced by the failure transitions.
Our improvements are strongest for WFSAs that are \emph{sparse} in a sense to be defined shortly.
We summarise our contributions as follows:
\begin{itemize}
    \item We present simple baseline algorithms using failure transition removal (\cref{sec:failure-expansion}) and memoization (\cref{sec:naive-improvement}).
    \item We present an algorithm for computing the pathsum of ring-weighted WFSAs, utilizing subtraction (\cref{sec:ring-backward}).
    \item With some extra work to avoid subtraction (\cref{sec:aggregator}), we extend the algorithm to general semirings (\cref{sec:general-backward}).
\end{itemize}

\section{Preliminaries} \label{sec:preliminaries}
This section defines WFSAs, the pathsum problem, the backward algorithm, and failure transitions.
\begin{defin}
A \defn{semiring} is a $5$-tuple $\semiring = \semiringtuple$ where $\semiringset$ is a set equipped with operations $\oplus$ and $\otimes$, s.t. $\additiontuple$ is a commutative monoid, $\multiplicationtuple$ is a monoid, $\oplus$ distributes over $\otimes$, and $\semizero$ annihilates $\otimes$.
\end{defin}

\begin{defin}
A \defn{weighted finite-state automaton} (WFSA) is a 5-tuple $\wfsa = \wfsatuple$, where $\alphabet$ is a finite alphabet, $\states$ a finite set of states, $\transitions$ a collection of transitions in $\states \times \alphabet \times \semiringset \times \states$, $\initf : \states \rightarrow \semiringset$ the initial-state weighting function, and $\finalf : \states \rightarrow \semiringset$ the final-state weighting function.
\end{defin}

To improve readability, we render a transition $\left(q, a, w, q'\right)$ as $\edge{q}{a}{w}{q'}$. 
We further define $\transitions(q) \defeq \{e \mid \exists a, w, \qchild: e = \edge{q}{a}{w}{\qchild} \in \transitions \}$ as the set of outgoing transitions of $q \in \states$, and $\transitions(q, a)$ as those labeled with $a \in \alphabet$.  
$\outsymbols(q) \defeq 
\{a \mid \transitions(q,a)\neq \emptyset\}$ 
denotes the set of transition labels in $\transitions(q)$.\looseness=-1

Importantly, we will assume that the graph $(\states,\transitions)$ is acyclic (see \cref{fn:cyclic}).  Less importantly,
our definition of WFSAs does not allow $\varepsilon$-transitions, assuming that they have been eliminated in advance \cite{mohri-2002-eps}, which is easy in the acyclic case.  Our runtime analyses assume for simplicity that 
\textit{(i)} the graph is connected (implying $|\transitions| \geq |\states| - 1$) and 
\textit{(ii)} that for each $q, q' \in \states$, $\transitions$ contains at most one transition $\edge{q}{a}{w}{q'}$ for any $a \in \alphabet$. This can always be achieved by replacing ``parallel'' transitions $\left\{\edge{q}{a}{w_i}{q'} \,\middle|\, i \right\} \subseteq \transitions$ with $\edge{q}{a}{\bigoplus_{i}w_i}{q'}$.

\begin{defin}
A \defn{path} $\apath$ in a WFSA $\wfsa$ is a sequence of consecutive transitions in $\transitions$, \\
\(
q_0 \xrightarrow{a_1 / w_1} q_1 \cdots q_{N-1} \xrightarrow{a_{N} / w_{N}} q_{N}. 
\)
$\prevq(\apath) \defeq q_0$ and $\nextq(\apath) \defeq q_{N}$ refer to the initial and final states of $\apath$, respectively.
$\paths(\wfsa)$ denotes the set of all paths in $\wfsa$.\looseness=-1
\end{defin}

\begin{defin}
The \defn{inner path weight} is defined as $\displaystyle \innerweight(\apath) \defeq \bigotimes_{n=1}^N w_n$ and the (full) \defn{path weight} as $\displaystyle \weight(\apath) \defeq\initf\left(\prevq(\apath)\right) \otimes \innerweight(\apath) \otimes \finalf\left(\nextq(\apath)\right)$. 
\end{defin}
\begin{defin}
The \defn{pathsum} of $\wfsa$ is defined as 
\begin{equation}
    \pathsum \left(\wfsa\right) \defeq \monsteroplus{\apath \in \paths \left( \wfsa \right)}  \weight(\apath) .
\end{equation}
\end{defin}
\noindent The problem of computing the pathsum is sometimes also referred to as the generalized shortest-distance problem \citep{10.5555/639508.639512}.
\begin{defin}
The \defn{backward value} $\backwardweight(q)$ of a state $q \in \states$ is the sum of the inner weights of all paths $\apath$ starting at $q$ right-multiplied by $\finalf\left(\nextq\left(\apath\right)\right)$, i.e., 
\begin{equation} \label{eq:backwardweight-state}
    \backwardweight(q) \defeq \monsteroplus{\substack{ \apath \in \paths(\wfsa), \\ \prevq(\apath) = q}}\innerweight(\apath) \otimes \finalf(\nextq (\apath) ).
\end{equation}
We extend this definition to state--symbol pairs $\left(q, a \right) \in \states \times \alphabet$ as
\begin{equation} \label{eq:backwardweight-state-symbol}
    \backwardweight(q, a) \defeq \monsteroplus{\edge{q}{a}{w}{\qchild} \in \transitions} w \otimes \backwardweight(\qchild).
\end{equation}
The value $\backwardweight(q, a)$ can be seen as the result of restricting the paths contributing to $\backwardweight(q)$ to those starting with $a \in \alphabet$.
\end{defin}
For $S \subseteq \alphabet$ we also define
\begin{equation}
    \label{eq:backward-symbol-subset}
    \backwardweight(q, S) \defeq \monsteroplus{a \in S} \backwardweight(q, a).
\end{equation}
Notice that $\backwardweight(q) = \finalf\left(q\right) \oplus \backwardweight(q, \outsymbols(q))$.

Na{\"i}vely computing the pathsum by enumerating all $\apath \in \paths (\wfsa)$ 
in an acyclic WFSA would result in an exponential runtime.
However, algebraic properties of semirings allow for faster algorithms \citep{10.5555/639508.639512}.
An example is the backward algorithm, a dynamic program which computes backward values and the pathsum in acyclic WFSAs in time $\bigO{}\left(|\transitions|\right)$.
It exploits the fact that, in acyclic WFSAs, $\states$ can always be topologically sorted and the backward values can be computed in \emph{reverse} topological order.
This guarantees that the backward values of $q$'s children will have been computed by the time we expand $q$, meaning that $\backwardweight(q)$ can be computed as \looseness=-1
\begin{equation} \label{eq:backward-recursion}
    \backwardweight(q) \gets \finalf(q) \oplus \monsteroplus[r]{\edge{q}{a}{w}{\qchild} \in \transitions} w \otimes \backwardweight(\qchild)
\end{equation}
The pseudocode is given in \cref{alg:backward}.
All our algorithms are based on the backward algorithm.\looseness=-1
\begin{algorithm}[h]
\caption{
}

\label{alg:backward}
\begin{algorithmic}[1]
\Func{$\backward(\wfsa)$}
\For{$q \in \reversetopo (\wfsa)$}
    \State $ \localterm{\backwardweight(q, \outsymbols(q))} \gets \displaystyle \monsteroplus[r]{\edge{q}{a}{w}{\qchild} \in \transitions} w \otimes \backwardweight{}(q')$ \label{line:bwd-standard-computation}
    \State $\backwardweight(q) \gets \finalf(q) \oplus \backwardweight(q, \outsymbols(q))$
\EndFor
\State \Return $\displaystyle \bigoplus_{q \in \states} \initf(q) \otimes \backwardweight(q)$ \InlineComment{Equals $\pathsum(\wfsa)$}
\EndFunc
\end{algorithmic}
\end{algorithm}

\subsection{Failure Transitions}
We consider an extension of WFSAs where any state can have a single \emph{fallback state} $\fallbackstate$. \looseness=-1
\begin{defin}
A WFSA with failure transitions (WFSA-$\phi$) is a 6-tuple $\wfsa = \failurewfsatuple$, where $\failuref$ is a \defn{failure function}---a partial function that maps some states $q \in \states$ to their \defn{fallback state} $\failuref(q) = \fallbackstate$.\looseness=-1
\end{defin}
Fallback states can be represented by transitions $\edge{q}{\phi}{\semione}{\fallbackstate}$ with a special meaning:\footnote{We use $\phi$ to refer to both the function and the symbol.} they are only traversed upon reading a symbol $a \notin \outsymbols(q)$ and thus represent a default option used when no ordinary transition is available.\footnote{\label{fn:extensions}A $\phi$-transition from a non-final state could be used at the end of the input string, to try to transition to a final state and accept the string.  But this case does not arise in our WFSA formalization, where every state $q$ is final with an explicit final-state weight  $\finalf(q)$ (possibly $\semizero$).  
Separately, if we were to extend our WFSA formalization to allow $\varepsilon$-transitions, then a $\phi$-transition could never be used at a state that also had an $\varepsilon$-transition, since the $\varepsilon$-transition would always be available.}
This formalization means that every state has at most one fallback state.  

We do not include $\phi$ in $\alphabet$ or $\phi$-transitions in $\transitions$. 
We denote the set of $\phi$-transitions as $\transitions^\phi$ and assume that $\transitions \cup \transitions^\phi$ still forms an acyclic graph. 

$\phi$-transitions can be explicitly represented in a normal WFSA by \emph{expansion} of $\phi$-transitions.
\begin{defin}
Given an acyclic WFSA-$\phi$ $\wfsa = \failurewfsatuple$, we introduce the recursively defined \defn{failure-expanded transition set} as follows
\begin{equation}
    \failureexpansion(q, a) \defeq \begin{cases}
\transitions(q, a) & \textbf{if } a \in \outsymbols(q)\\
\failureexpansion(\fallbackstate, a) & \textbf{if } q \text{ has a } \phi \text{ arc}\\
\emptyset & \textbf{otherwise}
\end{cases}
\end{equation}
and the set $\failureexpansion \subseteq \states \times \alphabet \times \semiringset \times \states$ as the union of these sets over $\states$ and $\alphabet$.
\end{defin}
$\failureexpansion(q, a)$ is well-defined due to the assumed acyclicity of $\transitions^\phi$.  It may be empty.
$\failureexpansion$ captures all ``indirect'' transitions which can be made across arbitrarily long paths of only $\phi$-transitions.
$\extendedoutsymbols(q)$, analogously to $\outsymbols(q)$, denotes the set of outgoing symbols for $q \in \states$ in the failure-expanded WFSA-$\phi$.\looseness=-1

\begin{defin} \label{def:avgoutfrac} 
We define the \defn{average out-symbol fraction} $\avgoutfrac$ of a WFSA as
\begin{equation}
    \avgoutfrac = \mean_{q\in \states} \frac{|\outsymbols(q)|}{|\alphabet|} = \frac{\sum_{q \in \states}|\outsymbols(q)|}{|\states||\alphabet|}.
\end{equation}
$\avgoutfrac \in [0, 1]$ is a measure of completeness of the WFSA. We correspondingly define $\overline{\avgoutfrac}$, the equivalent in the failure-expanded transition set
$\failureexpansion$.\end{defin}
We say informally
that a WFSA is \defn{$\alphabet$-sparse} if $\avgoutfrac \ll 1$, so on average $|\outsymbols(q)|
\ll |\alphabet|$.
Intuitively, this means that the average state only has outgoing transitions on a few distinct symbols. We will show that the runtime tradeoff between our baseline pathsum algorithm  $\memoizationbackward$ (\cref{alg:aggregation-backward}) and later algorithms depends on the \emph{difference} between $s$ and $\overline{\avgoutfrac}$. 
Our algorithms are efficient when $\avgoutfrac \ll \overline{\avgoutfrac}$: intuitively in the regime where failure expansion would add outgoing transitions for many new symbols.\looseness=-1

Correcting \cref{eq:backward-recursion} to take $\phi$-transitions into account, the backward values in a WFSA-$\phi$ can be computed as
\begin{equation} \label{eq:backward-phi-recursion}
    \backwardweight(q) \gets
     \finalf(q) \oplus \monsteroplus[r]{\edge{q}{a}{w}{\qchild} \in \failureexpansion} w \otimes \backwardweight(\qchild).
\end{equation}
Importantly, the following equality holds 
\begin{equation}  \label{eq:backwardweight-phi-state-symbol-decomposition}
    \backwardweight(q, a) = \begin{cases} 
        {\displaystyle \quad\ \monsteroplus{\edge{q}{a}{w}{\qchild} \in \transitions}\ w \otimes \backwardweight(\qchild)} & \textbf{ if } a \in \outsymbols(q) \\
        \backwardweight(\fallbackstate, a) & \textbf{ otherwise }
    \end{cases}
\end{equation}
This follows straight from the definition of $\failuref$.
It states that the backward values of the state-symbol pairs $\left(q, a\right)$ in WFSA-$\phi$ equal the ones in a normal WFSA if an $a$-labeled transition can be taken; if not, the backward value is inherited from the fallback state, since the $\semione$-weighted $\phi$-transition is taken.

Connected components of the graph formed by $\phi$-transitions of a WFSA-$\phi$ are \emph{trees} (specifically, anti-arborescences) since a state can have at most one outgoing $\phi$-transition and the WFSA is acyclic.
This motivates the following definition.
\begin{defin} 
Let $\wfsa$ be an acyclic WFSA-$\phi$. 
A \defn{failure tree} $\tree$ is a connected component of the graph formed by $\phi$-transitions of $\wfsa$.
\end{defin}
\noindent An example of a failure tree $\tree$ is shown in \cref{fig:failure-tree}.
We write $|\tree|$ for the number of states in $\tree$, with $\maxfailuretreesize$ being the number of states in the largest failure tree and $\maxfailurepathsize$ the number of states in the longest failure path.

$\tree_q$ denotes the failure tree containing $q \in \states$.  We write $q \prec q'$ to say that $q$ is a proper ancestor of $q'$ in $\tree_q$, i.e., there is a non-empty $\phi$-path from $q$ to $q'$.

\section{Expanding Failure Transitions}

The pathsum of a WFSA-$\phi$ can be na{\"i}vely computed by replacing the $\phi$-transitions with normal ones according to the semantics of the $\phi$-transitions and running the backward algorithm on the expanded WFSA. 
Before introducing our contributions, we present this method for pedagogical purposes.
While this solution is near-optimal for non-$\alphabet$-sparse WFSAs, it can be improved for certain $\alphabet$-sparse WFSAs.   



\subsection{Expanding Failure Transitions} \label{sec:failure-expansion}
\defn{Failure expansion} is a transformation of an acyclic WFSA-$\phi$ which replaces the $\phi$-transitions while retaining acyclicity.
See \cref{alg:failure-expansion} for the pseudocode, \cref{fig:phi-expansion} for an example of failure expansion, and \cref{appendix:algo-walkthrough} for an example of how the backward algorithm operates in this setting. \looseness=-1
\begin{algorithm}[h]
\caption{
}
\label{alg:failure-expansion}
\begin{algorithmic}[1]

\Func{$\failureexpansionprocedure(\wfsa)$}

\State $\failureexpansion \gets \transitions$ \InlineComment{Will be updated}

\For{$q \in \reversetopo (\transitions^\phi)$}
\State $\failureexpansion \gets \failureexpansion 
\cup \{ \edge{q}{a}{w}{q'} \mid $
\State \hfill $\edge{\fallbackstate}{a}{w}{q'} \in \failureexpansion, a \notin \outsymbols(q) \}$
\EndFor
\State \Return $\failureexpansion$
\EndFunc
\end{algorithmic}
\end{algorithm}
\begin{figure}
    \centering
    \begin{subfigure}[b]{\linewidth}
        \centering
        \begin{tikzpicture}[node distance = 8 mm]
        \footnotesize
        \node[state, initial] (q1) [] { 1 }; 
        \node[state] (q2) [above right = of q1] { 2 }; 
        \node[state] (q3) [below = 10mm of q2] { 3 }; 
        \node[state] (q4) [right = of q2] { 4 }; 
        \node[state] (q5) [right = of q3] { 5 }; 
        \node[state, accepting] (q6) [below right = of q4] { 6 }; 
        \node[state] (q7) [above right = of q4] {7};
        \node[state] (q8) [left = of q7] {8};
        \draw[-{Latex[length=2mm]}] 
        (q1) edge[above left, swap, phitransitioncolor] node{ $\phi$ } (q2) 
        (q2) edge[above, swap, phitransitioncolor] node{ $\phi$ } (q4) 
        (q2) edge[above left, swap] node{ $a$ } (q3) 
        (q4) edge[above right, swap] node{ $a$ } (q6) 
        (q4) edge[above left, swap] node{ $b$ } (q5)
        (q7) edge[above left, swap, phitransitioncolor] node{ $\phi$ } (q4)
        (q7) edge[above, swap] node{ $b$ } (q8);
        \end{tikzpicture}
        \caption{Example of a {\color{phitransitioncolor} failure tree}. Its root is node $4$.}
        \label{fig:failure-tree}
    \end{subfigure}
    \par\bigskip\medskip
    \begin{subfigure}[b]{\linewidth}
        \centering
        \begin{tikzpicture}[node distance = 8 mm]
        \footnotesize
        \node[state, initial] (q1) [] { 1 }; 
        \node[state] (q2) [above right = of q1] { 2 }; 
        \node[state] (q3) [below right = of q1] { 3 }; 
        \node[state] (q4) [right = of q2] { 4 }; 
        \node[state] (q5) [right = of q3] { 5 }; 
        \node[state, accepting] (q6) [below right = of q4] { 6 }; 
        \node[state] (q7) [above right = of q4] {7};
        \node[state] (q8) [left = of q7] {8};
        \draw[-{Latex[length=2mm]}] 
        (q1) edge[above left, swap, dashed, phitransitioncolor] node{ $\phi$ } (q2) 
        (q2) edge[above, swap, dashed, phitransitioncolor] node{ $\phi$ } (q4) 
        (q2) edge[above left, swap] node{ $a$ } (q3) 
        (q4) edge[above right, swap] node{ $a$ } (q6) 
        (q4) edge[above left, swap] node{ $b$ } (q5)
        (q7) edge[above left, swap, dashed, phitransitioncolor] node{ $\phi$ } (q4)
        (q7) edge[above, swap] node{ $b$ } (q8)
        
        (q1) edge[below left, swap, failureexpansioncolor, dashed] node{ $a$ } (q3)
        (q1) edge[above right, swap, failureexpansioncolor, dashed] node{ $b$ } (q5)
        (q7) edge[right, swap, failureexpansioncolor, dashed] node{ $a$ } (q6)
        (q2) edge[below left, swap, failureexpansioncolor, dashed] node{ $b$ } (q5);
        \end{tikzpicture}
        \caption{To expand failure transitions, the {\color{failureexpansioncolor} dashed transitions} are added and the {\color{phitransitioncolor} $\phi$-transitions} are removed.}
        \label{fig:phi-expansion}
    \end{subfigure}
\end{figure}

In deterministic WFSA-$\phi$'s, failure expansion adds $\bigO{\left(|\states||\alphabet|\right)}$ new transitions, since each of the $|\states|$ states $q$ gains new transitions to each of $\fallbackstate$'s children. 
More precisely, $q$ gains up to $| \overline{\outsymbols}(q) \setminus \outsymbols(q)|$ transitions,
where $\overline{\outsymbols}(q) \defeq \{a: \failureexpansion(\fallbackstate, a) \neq \emptyset \}$ denotes the set of symbols on the outgoing transitions from $\fallbackstate$ in the failure-expanded automaton.
In non-deterministic WFSA-$\phi$'s, however, $q$ can gain up to $|\states| | \overline{\outsymbols}(q) \setminus \outsymbols(q)|$ transitions, since $\fallbackstate$ might have up to $|\states|$ $a$-labeled transitions $\forall a \in \alphabet$.
This results in $\bigO{\left(|\states|^2|\alphabet|\right)}$ new transitions.
Following the derivation in \cref{appendix:failure-expansion-num-added-transitions}, the runtime of the backward algorithm on the $\phi$-expanded WFSA-$\phi$ is therefore $\bigO(|\failureexpansion|) = \bigO{(|\transitions| + |\states|^2 (\overline{\avgoutfrac} - \avgoutfrac) |\alphabet|)}$ in the general case and $\bigO{(\failureexpansion)} = \bigO{(|\transitions| + |\states| (\overline{\avgoutfrac} - \avgoutfrac) |\alphabet|)}$ in deterministic WFSAs.\looseness=-1

\subsection{Decomposing the Backward Values} \label{sec:naive-improvement}

The algorithms we present in later sections sidestep the need to materialize all additional transitions replacing the failure transition.
They are based on a decomposition of the backward values into two components: the local and the failure component.
Using \cref{eq:backward-symbol-subset}, we can split $\backwardweight(q)$ into
\begin{align} \label{eq:backward-failure-recursion}
    \backwardweight(q) &= \finalf(q) \oplus \backwardweight(q, \alphabet) \\
    \label{eq:backward-split-outgoing-and-rest}
        \backwardweight(q, \alphabet)
    &= 
        \localterm{\underbrace{\backwardweight(q, \outsymbols(q))}_{\text{local}}}
        \oplus 
        \failureterm{\underbrace{\backwardweight(q, \alphabet \setminus \outsymbols(q))}_{\text{failure}}}.
\end{align}
The two terms on the right hand-side of \cref{eq:backward-split-outgoing-and-rest} can be further expanded as
\begin{align}
\label{eq:backward-split-outgoing}
    \localterm{\backwardweight\left(q, \outsymbols(q)\right)}    
    &= \monsteroplus[r]{\edge{q}{a}{w}{q'} \in E} w \otimes \backwardweight(q')
\\
    \label{eq:backward-sigma-sum}
        \failureterm{\backwardweight(q, \alphabet \setminus \outsymbols(q))}
    &= \monsteroplus[r]{b \in \alphabet \setminus \outsymbols(q)} \backwardweight(\fallbackstate, b)
\end{align}
except that the second term is $\semizero$ if $q$ has no failure transition (in which case $\fallbackstate$ is not defined). 
$\localterm{\backwardweight(q, \alphabet(q))}$ is exactly the quantity computed by \cref{alg:backward} on \cref{line:bwd-standard-computation}; our modifications never change this computation.
Rather, all of our algorithms seek to simplify the computation of $\failureterm{\backwardweight(q, \alphabet \setminus \outsymbols(q))}$.

\cref{eq:backward-sigma-sum} makes it possible to avoid failure expansion by storing not only $\backwardweight(q)$ but also the values $\backwardweight(q, a)$ at each state $q$.  
Since $\fallbackstate$ will then memoize all needed $\backwardweight(\fallbackstate, b)$ values, the sum \labelcref{eq:backward-sigma-sum} becomes easy to compute for any $q$ that may back off to $\fallbackstate$.  
Passing the summand $\backwardweight(\fallbackstate, b)$ back to $q$ is cheaper than 
passing back all of the arcs $\edge{\fallbackstate}{b}{w}{\qchild} \in \failureexpansion$ that contribute to that summand, as \cref{alg:failure-expansion} does: a non-deterministic WFSA may have multiple such arcs.
The pseudocode of this modification is presented in \cref{alg:aggregation-backward}.
Notice the additional term $\failureterm{\backwardweight(q, \alphabet \setminus \outsymbols(q))}$ on \cref{line:aggregation-backward-additional-term} in \cref{alg:aggregation-backward}, which was not needed in the backward algorithm for ordinary WFSAs.
See \cref{appendix:algo-walkthrough} for a guided example on a small WFSA. \looseness=-1

In the general case of non-deterministic WFSAs, 
failure expansion may have to loop over as many as $|\states||\alphabet \setminus \outsymbols(q)|$ transitions at each state $q$. \Cref{alg:aggregation-backward} reduces this to a loop over $|\alphabet \setminus \outsymbols(q)|$ symbols, which is $(\overline{\avgoutfrac} - \avgoutfrac) |\alphabet|$ on average.  The full complexity of \cref{alg:aggregation-backward} is then $\bigO{(|\transitions| + (\overline{\avgoutfrac} - \avgoutfrac) |\alphabet| |\states|))}$ (similarly to \cref{appendix:failure-expansion-num-added-transitions}).

The shortcoming of \cref{alg:aggregation-backward} is that $(\overline{\avgoutfrac} - \avgoutfrac) |\alphabet|$ may still be large.  The terms $\backwardweight(\fallbackstate, b)$ must be individually copied back to $q$ as $\backwardweight(q, b)$ for each $|\alphabet \setminus \outsymbols(q)|$.
Our proposed algorithms in the following subsections avoid the overhead incurred by this copying.

\begin{algorithm}[h]
\caption{
}
\label{alg:aggregation-backward}
\begin{algorithmic}[1]
\Func{$\memoizationbackward(\wfsa)$}

\For{$q \in \reversetopo (\wfsa)$}

\For{$a \in \outsymbols(q)$}
\State $\localterm{\backwardweight(q, a)} \gets \monsteroplus{\edge{q}{a}{w}{\qchild} \in \transitions} w \otimes \backwardweight(\qchild)$
\EndFor

\State $\failureterm{\backwardweight(q,\alphabet \setminus \outsymbols(q))} \gets \semizero$
\If{$q$ has a fallback state}
\For{$b \in \alphabet \setminus \outsymbols(q)$}
\State $\backwardweight(q, b) \gets \backwardweight(\fallbackstate, b)$ \label{line:aggregation-backward-failure-copy}
\State $\failureterm{\backwardweight(q,\alphabet \setminus \outsymbols(q))} \opluseq \backwardweight(\fallbackstate, b)$\label{line:aggregation-backward-failure-sum}
\EndFor
\EndIf
\State $\backwardweight(q, \alphabet) \gets \localterm{\backwardweight(q, \outsymbols(q))} \oplus \failureterm{\backwardweight(q,\alphabet \setminus \outsymbols(q))}\!\!$ \label{line:aggregation-backward-additional-term}
\State $\backwardweight(q) \gets \finalf(q) \oplus \backwardweight(q,\alphabet) $
\EndFor
\State \Return $\bigoplus_{q \in \states} \initf(q) \otimes \backwardweight(q)$
\EndFunc
\end{algorithmic}
\end{algorithm}

\section{An Algorithm with Subtraction} \label{sec:ring-backward}


\cref{alg:aggregation-backward} computes $\backwardweight(q)$ in part by summing up to $|\alphabet \setminus \outsymbols(q)|$ values passed back from $\fallbackstate$.
This section presents a more efficient algorithm for \emph{ring}-weighted $\alphabet$-sparse WFSAs. 
As rings allow \emph{subtraction}, we can compute the failure term
as follows:\looseness=-1
\begin{equation} \label{eq:backward-minus}
    \failureterm{\backwardweight(q, \alphabet \setminus \outsymbols(q))} = \backwardweight(\fallbackstate, \alphabet) \ominus \backwardweight(\fallbackstate,\outsymbols(q))
\end{equation}
Recall that $\beta(\fallbackstate,\outsymbols(q)) \defeq \bigoplus_{a
  \in \outsymbols(q)} \backwardweight(\fallbackstate, a)$ by \labelcref{eq:backward-symbol-subset}. Thus \cref{eq:backward-minus} effectively uses $|\outsymbols(q)|$ subtractions (for $a \in \outsymbols(q)$), whereas \cref{eq:backward-sigma-sum} used $|\alphabet \setminus \outsymbols(q)|$ additions (for $b \in \alphabet \setminus \outsymbols(q)$).  In the runtime analysis, these subtractions are already covered by the $\bigO\left(|\outsymbols(q)|\right)$ runtime needed for the $|\outsymbols(q)|$ additions in \cref{eq:backward-split-outgoing}.  Overall, \cref{eq:backward-split-outgoing-and-rest,eq:backward-split-outgoing,eq:backward-minus} compute $\backwardweight(q, \alphabet)$ in \cref{eq:backward-split-outgoing-and-rest} by combining $\oplus$ and $\ominus$ to \emph{replace} just $|\outsymbols(q)|$ of the summands of $\backwardweight(\fallbackstate, \alphabet)$---namely, those overridden at $q$.\looseness=-1

But how fast is it to \emph{find} the subtrahends $\backwardweight(\fallbackstate, a)$ for $a \in \outsymbols(q)$?  Eagerly storing $\backwardweight(q,a)$ (if non-$\semizero$) for \emph{every} $q\in\states,a\in\alphabet$ (in case it is needed during backoff) would allow constant-time lookup, but doing so
would require copying $\backwardweight(\fallbackstate, b)$ backward to $\backwardweight(q, b)$ for all $b \in \alphabet \setminus \outsymbols(q)$, just as in \cref{alg:aggregation-backward}, which would incur the same complexity of $\bigO{(|\alphabet \setminus \outsymbols(q)|)}$.  
So instead of computing and storing the full set of $\backwardweight(\fallbackstate, a)$ values $\forall a \in \alphabet$, we will compute on demand only the ones that need replacement.  This involves following $\phi$-arcs forward until we find an $a$ arc, or run out of $\phi$ arcs, or encounter a memo because $\backwardweight(\fallbackstate, a)$ was already needed by a different ancestor of $\fallbackstate$ in its failure tree.
The full algorithm is presented as \cref{alg:ring-backward}.\footnote{\label{fn:toporingbackward}In practice, it is handy for the summation in $\ringbackward$ \cref{line:ring-backward-sum} to visit
  the states $\states$ in reverse topological order, as was done explicitly in \cref{alg:aggregation-backward}.  This slightly simplifies implementation of $\backwardweight(q,\alphabet)$ since whenever that function is called,
  the call is guaranteed to be memoized already.  But this guarantee cannot apply to the $\backwardweight(q,a)$ function, as it is called only for those $a$ where it is needed.  In particular, when $\backwardweight(\fallbackstate,a)$  is called in \cref{alg:ring-backward}, the memo (for that particular $a$) may not yet exist and will have to be filled in on demand.}



\begin{algorithm}[h]
\caption{
}
\label{alg:ring-backward}
\begin{algorithmic}[1]
\Func{$\ringbackward(\wfsa)$}
\State \Return $\displaystyle \bigoplus_{q \in \states} \initf(q) \otimes \backwardweight(q)$\label{line:ring-backward-sum}
\EndFunc
\Func{$\backwardweight(q)$}  
\State \Return $\backwardweight(q) \gets \finalf(q) \oplus \backwardweight(q, \alphabet)$
\EndFunc
\vspace{1mm}
\Func{$\backwardweight(q, \alphabet)$} \InlineComment{Memoizes its result}
\State $\localterm{\backwardweight(q, \outsymbols(q))} \gets \bigoplus_{a \in \outsymbols(q)} \backwardweight(q, a)$\label{line:ring-plus}
\If{$q$ has no fallback state}
\Return $\localterm{\backwardweight(q, \outsymbols(q))}$
\EndIf
\State $\backwardweight(\fallbackstate, \outsymbols(q)) \gets \bigoplus_{a \in \outsymbols(q)} \backwardweight(\fallbackstate, a)$\label{line:ring-minus}
\Statepar{\Return \\ \hspace*{2ex}$\failureterm{\backwardweight(\fallbackstate,\alphabet)} \oplus \big( \localterm{\backwardweight(q, \outsymbols(q))}\failureterm{\mbox{}\ominus \backwardweight(\fallbackstate, \outsymbols(q))}\big)$}\label{line:ring-backward-replace}
\EndFunc
\vspace{1mm}
\Func{$\backwardweight(q, a)$} \InlineComment{Memoizes its result}\label{line:memoize-qa}
\If{$a \in \outsymbols(q)$} \Return $\monsteroplus{\edge{q}{a}{w}{\qchild} \in \transitions} w \otimes \backwardweight(\qchild)$\label{line:ring-backward-out-arcs}
\ElsIf{$q$ \scalebox{0.95}{has a fallback state}} \Return $\backwardweight(\fallbackstate,a)$\label{line:ring-recurse}
\Else\ \Return $\semizero$
\EndIf
\EndFunc
\end{algorithmic}
\end{algorithm}

\subsection{Runtime}\label{sec:ring-backward-runtime}

The runtime of \cref{alg:ring-backward} is on the order of the number of calls to \cref{line:memoize-qa}, plus $|E|$ to cover all the sums in \cref{line:ring-backward-out-arcs} (which executes at most once for each $q,a$ pair, thanks to memoization).  Every $a \in \outsymbols(q)$ results in two such calls, at \cref{line:ring-plus,line:ring-minus}; there is also a possible recursive call at \cref{line:ring-recurse} if $a \in \outsymbols(\ancstate)$ for at least one proper ancestor $\ancstate \prec q$ in the failure tree (thanks to memoization, this happens at most once per $q,a$ pair, even if there are multiple choices of $\ancstate$).
Thus, the overall runtime is $\bigO \left(|\transitions| + \sum\limits_{q \in \states} |\hat{\outsymbols}(q)| \right)$, where $\hat{\outsymbols}(q) \subseteq \alphabet$ is defined as $\bigcup_{\ancstate \preceq q} \outsymbols(\ancstate)$.
A looser bound written in terms of $\avgoutfrac$ is $\bigO{\left(|\transitions| + |\alphabet| |\states| \min(1,s\maxfailurepathsize) \right)}$.\footnote{Clearly $\sum_{q \in \states} |\hat{\outsymbols}(q)| \leq |\alphabet| |\states|$ since $|\hat{\outsymbols}(q)| \leq |\alphabet|$.  Also, 
$\sum_{q \in \states} |\hat{\outsymbols}(q)| 
\leq \sum_{q \in \states} \sum_{\ancstate \preceq q} |\outsymbols(\ancstate)| 
= \sum_{\ancstate \in \states} \sum_{q \succeq \ancstate} |\outsymbols(\ancstate)| 
\leq \sum_{\ancstate \in \states}\! \maxfailurepathsize |\outsymbols(\ancstate)| 
=\!\maxfailurepathsize\! \sum_{\ancstate \in \states}\!|\outsymbols(\ancstate)| 
= \maxfailurepathsize \avgoutfrac|\alphabet||\states|$.\looseness=-1}

We will revisit ring-weighted WFSA-$\phi$'s in \cref{sec:ring-aggregator}.

\section{Incrementally Modified Aggregator} \label{sec:aggregator}

The point of \cref{alg:ring-backward} \cref{line:ring-backward-replace} is to replace some summands of $\backwardweight(\fallbackstate,\alphabet)$ to get $\backwardweight(q,\alphabet)$.  When no subtraction operator $\ominus$ is available (e.g., if $\oplus=\max$), we can use an \defn{aggregation data structure} that is designed to efficiently replace individual summands in a sum without using subtraction.  For example, a \defn{Fenwick tree} \citep{fenwick1994new}
can replace a summand and recompute the sum in $\bigO(\log N)$ time, where $N$
is the number of summands.  (Fenwick trees are similar to binary
heaps; they are reviewed in
\cref{appendix:fenwick}.) Here we merely give the interface
to aggregators:

\begin{algorithm}[h]
\begin{algorithmic}[1]

\State \textbf{class} \aggregatortype{}(): \InlineComment{We use $\aggregator$ to refer to an aggregator instance}

\State \quad\textbf{def} \myfunc{set}($a$: $\alphabet$, $v$: $\semiringset$) \InlineComment{Updates $\aggregator(a) \gets v$}
\State \quad\textbf{def} \myfunc{get}($a$: $\alphabet$) $\rightarrow \semiringset$ \InlineComment{Returns $\aggregator(a)$ (default $\semizero$)}
\State \quad\textbf{def} \myfunc{value}() $\rightarrow \semiringset$ 
\InlineComment{Returns $\displaystyle \smashoperator[r]{\bigoplus_{a \in \alphabet}} \myfunc{get} (a)$ }
\State \quad\textbf{def} \myfunc{undo}($n$: $\mathbb{N})$ 
\InlineComment{Reverts the last $n$ updates} 
\alglinenumNew{aggregator}
\alglinenumSave{aggregator}
\end{algorithmic}
\end{algorithm}

We will represent each sum $\backwardweight(q,\alphabet)$ in \cref{alg:ring-backward} as the total value of an aggregator that stores summands $\backwardweight(q,a)$ for $a \in \alphabet$.  In principle, this aggregator could be obtained by copying the aggregator for $\backwardweight(\fallbackstate,\alphabet)$ and then modifying some summands (see \cref{line:ring-backward-replace}). However, aggregators are not constant-size data structures, so creating all of these slightly different aggregators would be expensive.  

Instead, our strategy will be to use just a \emph{single} aggregator, for the ``current'' state $q$, and make small modifications as we visit different states $q'$.  More precisely, we have one aggregator $\aggregator$ per failure tree, first created at the tree's root.  When we step backwards in the failure tree, say from $\fallbackstate$ to $q$, we modify ``just a few'' summands in $\aggregator$ so that $\backwardweight(q,a)$ replaces $\backwardweight(\fallbackstate,a)$ for $a \in \outsymbols(q)$.  This is fast if $\outsymbols(q)$ is small.  We can now obtain $\backwardweight(q,\alphabet)$ as the aggregator's new total value.  To visit other ancestors of $\fallbackstate$, we must first move forward to $\fallbackstate$ again, which we do by reverting the modifications.%
\footnote{\label{fn:undo}Any data structure can support a $\myfunc{undo}$ method.  Any update method  begins by pushing a sentinel onto an \emph{undo stack}; whenever it writes a new value to a memory cell, it pushes an operation that would write the old value to that memory cell.  $\myfunc{undo}(n)$ simply pops and applies these operations in reverse order, until it has popped $n$ sentinels.  Reverting updates in this way takes no more time than the original updates.}



\begin{defin}\label{def:represent}
Aggregator $\aggregator$ \defn{represents}
$q \in \states$ if
\begin{equation}
    \aggregator(a) = \backwardweight(q, a),\; \forall a \in \alphabet \nonumber
\end{equation}
\end{defin}
\noindent $\aggregator$ will be updated to represent different states in the failure tree at different times.  When $\aggregator$ represents $q$, it holds that $\backwardweight(q) = \finalf(q) \oplus \aggregator.\myfunc{value}()$, by \labelcref{eq:backward-failure-recursion}.

Updates are carried out by the methods in \cref{alg:update-restore}, which move backward and forward in a failure tree.  When $\aggregator$ represents $\fallbackstate$, we can call $\updatestep(\aggregator, q)$ to update $\aggregator$ so that it represents $q$.  At any later time when $\aggregator$ again represents $q$, we can call $\restorestep(\aggregator, q)$ to undo this update, so that $\aggregator$ again represents $\fallbackstate$.

\begin{algorithm}[h]
\caption{
}
\label{alg:update-restore}
\begin{algorithmic}[1]

\Func{$\updatestep(\aggregator, q)$}
\InlineComment{update $\aggregator$ that represented $\fallbackstate$ to represent $q$}
\For{$a \in \outsymbols(q)$}\label{line:visit-override}
    \State $\displaystyle \aggregator.\myfunc{set}(a, \backwardweight(q, a))$%
    \InlineComment{Use the memoizing $\backwardweight(q,a)$ from \cref{alg:ring-backward}}
\EndFor
\EndFunc


\Func{$\restorestep(\aggregator, q)$}
\InlineComment{update $\aggregator$ that represented $q$ to represent $\fallbackstate$}
\State $\myfunc{undo}(|\outsymbols(q)|)$ \InlineComment{revert all the updates made by $\updatestep$}
\EndFunc
\end{algorithmic}
\end{algorithm}

Each $\updatestep(\aggregator, q)$ or $\restorestep(\aggregator, q)$ call runs in time $\bigO(|\outsymbols(q)| \log |\alphabet|)$
, since it sets $|\outsymbols(q)|$ values in $\aggregator$.\footnote{A tighter analysis allows us to reduce this runtime to $\bigO(\log \max_{q \in \states} |\extendedoutsymbols(q)|)$; see \cref{appendix:fenwick} for details.}
Note that $\updatestep(\aggregator, q)$ accomplishes the same goal as \cref{alg:ring-backward} \cref{line:ring-backward-replace}, but with an extra runtime factor of $\bigO(\log |\alphabet|)$ to avoid subtraction.
It may still be faster than the $\bigO(|\alphabet \setminus \outsymbols(q)|)$ runtime of \cref{alg:aggregation-backward} \cref{line:aggregation-backward-additional-term}, when $|\outsymbols(q)|$ is quite small relative to $|\alphabet|$.

\section{A General Backward Algorithm for Acyclic WFSAs with Failure Transitions} 
\label{sec:general-backward}

\Cref{alg:new-backward-general} is our most general version of the backward algorithm for computing the pathsum of an acyclic WFSA-$\phi$.  It makes use of the \aggregatortype{} and pseudocode from the previous section.

\begin{algorithm}[h]
\caption{
}
\label{alg:new-backward-general}
\begin{algorithmic}[1]

\Func{$\genericbackward(\wfsa)$}

\For{$q \in \reversetopo (\wfsa)$}\label{line:general-topo}
  \State $\tree \gets \tree_q$ \InlineComment{failure tree containing $q$}\label{line:choose-tree}
  \If{$q$ has no fallback state} \InlineComment{$q$ is root of failure tree}
    \State $\aggregator_\tree \gets $ \textbf{new} \myfunc{Aggregator}() \InlineComment{New empty aggregator}
    \State $\updatestep(\aggregator_\tree, q)$; $q_\tree \gets q$ \InlineComment{Initialize $\aggregator_\tree$ \& remember $q$}\label{line:visit-root}
  \Else
  \While{$q_\tree$ is not a descendant of $q$ in $\tree$}\label{line:check-descendant}
    \State{$\restorestep(\aggregator_\tree, q_\tree)$; $q_\tree \gets q_\tree^\phi$} \InlineComment{Descend in $\tree$}
  \EndWhile
  \State{$\updatesteps(\aggregator_\tree, q, q_\tree)$; $q_\tree \gets q$} \InlineComment{Ascend in $\tree$}
  \EndIf
  \LineComment{Now $\aggregator_\tree$ represents $q$ (thanks to all of the above)}
  \State $\backwardweight(q) \gets \finalf(q) \oplus \aggregator_\tree.\myfunc{value}()$ \label{line:new-backward-general-failure}
\EndFor
\State \Return $\displaystyle \bigoplus_{q \in \states} \initf(q) \otimes \backwardweight(q)$ 
\EndFunc

\Func{$\updatesteps(\aggregator,q,q')$} \InlineComment{update $\aggregator$ that represented $q'$ to represent $q$}
   \If{$\fallbackstate \neq q'$}
       $\updatesteps(\aggregator,\fallbackstate,q')$; \label{line:visit-parent}
   \EndIf
   \State {$\updatestep(\aggregator,q)$}
\EndFunc
\end{algorithmic}
\end{algorithm}

Like \cref{alg:aggregation-backward}, this computes $\backwardweight(q)$ at all states in reverse topological order.  However, it attempts to share work among states $q$ in the same failure tree $\tree$, by having them share an aggregator $\aggregator_\tree$ that currently represents some state $q_\tree \in \tree$ (in the sense of \cref{def:represent}).  The algorithm updates the aggregator to represent $q$, by descending in the failure tree until it reaches a common descendant, and then ascending again until it reaches $q$.

To make \cref{line:check-descendant} efficient, we preprocess each failure tree by visiting its states in depth-first order and annotating each state with the time interval during which it is on the stack.\footnote{During depth-first search, the ``clock'' is a counter that starts at 0 and advances by 1 on each recursive call or return.  Thus, the ``time interval'' is a pair of small integers.}
The loop at \cref{line:check-descendant} continues until the $q_\tree$ interval contains the $q$ interval.

\subsection{Runtime}\label{sec:new-backward-general-runtime}

As in \cref{alg:ring-backward} (see \cref{sec:ring-backward-runtime}), $\bigO(|\transitions|)$ runtime is needed to sum over the non-failure transitions from each state.  The rest of the runtime is dominated by the calls to $\updatestep$ and $\restorestep$.  Recall from \cref{sec:aggregator} that visiting or leaving $q$ takes time $\bigO(|\outsymbols(q)| \log |\alphabet|)$.  Since a state can be left at most once for each time it is visited, it suffices to count just the visits.

The number of visits to each state depends on the (reverse) topological order used at \cref{line:general-topo}.  In the \emph{best case}, $q$ iterates over the states of each failure tree in depth-first order, starting at the root.  Then $\updatestep$ is called only on the current iterate $q$---either as a root (\cref{line:visit-root}) or as a parent (\cref{line:visit-parent}).  Since each state is $\updatestep$ed exactly once, the total runtime is
$\bigO(|\transitions|+\sum_{q \in Q} |\outsymbols(q)| \log |\alphabet|)$.
In the \emph{worst case}, however, each $q$ at \cref{line:general-topo} is far in the failure tree from the previous one, forcing $q_\tree$ to descend all the way to the root and then ascend again to $q$.
This means \cref{line:visit-parent} visits all states $q'$ for which $q \preceq q'$.  
The total runtime is therefore $\bigO(|\transitions| + \sum_{q' \in \states} |\outsymbols(q')|\,\ancs(q') \log |\alphabet|)$, where $\ancs(q') \defeq |\{q: q \preceq q'\}|$ is the number of ancestors of $q'$ in the failure tree.
Renaming the summation variable, we get $\bigO(|\transitions| + \sum_{q \in \states} |\outsymbols(q)|\,\ancs(q) \log |\alphabet|)$.\footnote{\label{fn:skiproot}This can be slightly improved by noting that \cref{line:visit-parent} never $\updatestep$s the root of the failure tree, so the $\sum_{q \in \states}$ can omit the root.  (Thus, the runtime is $\bigO(|\transitions|)$ on a WFSA without failure arcs.)
Although the root is still $\updatestep$ed once at \cref{line:visit-root}, the cost of that visit can be folded into the $|\transitions|$ term, since the initial creation of the aggregator at the root state $q$ can be accomplished in time only $\bigO(|\outsymbols(q)|)$ without the $\log$ factor (see \cref{appendix:fenwick}).}  

We can get a simpler but looser worst-case bound by increasing $\ancs(q)$ to $\maxfailuretreesize$, the maximum size of any failure tree.  Rewriting this in terms of $\avgoutfrac$, we have bounded the runtime by $\bigO ( |\transitions| + \avgoutfrac |\alphabet| |\states| \maxfailuretreesize  \log |\alphabet|)$, where, however, in the best case we avoid the $\maxfailuretreesize$ factor.

\begin{figure}
    \centering
    \begin{tikzpicture}[node distance = 6 mm]
    \footnotesize
    \node[state] (q1) [] { $1$ }; 
    \node[state] (q2) [right = of q1, yshift=8mm] { 2 }; 
    \node[state] (q3) [right = of q1, yshift=-8mm] { 3 };  
    \node[state] (q4) [right = of q2] { 4 }; 
    \node[state] (q5) [right = of q3] { 5 }; 
    \node (d1) [right = of q4] { $\ldots$ }; 
    \node (d2) [right = of q5] { $\ldots$ }; 
    \node[state] (q6) [right = of d1] { }; 
    \node[state] (q7) [right = of d2] { }; 
    \draw[-{Latex[length=2mm]}] 
    
    (q3) [left, black] edge node{ {\smaller $a$} } (q2) 
    (q4) [above left, black] edge node{ {\smaller $a$} } (q3) 
    (q5) [left, black] edge node{ {\smaller $a$} } (q4) 
    (d1) [above left, black] edge node{ {\smaller $a$} } (q5) ;
    
    \draw[-{Latex[length=2mm]}] 
    (q2) [above, phitransitioncolor] edge node{ \failureterm{\smaller $\phi$} } (q1) 
    (q3) [above, phitransitioncolor] edge node{ \failureterm{\smaller $\phi$} } (q1) 
    (q4) [above, phitransitioncolor] edge node{ \failureterm{\smaller $\phi$} } (q2) 
    (q5) [above, phitransitioncolor] edge node{ \failureterm{\smaller $\phi$} } (q3) 
    (d1) [above, phitransitioncolor] edge node{ \failureterm{\smaller $\phi$} } (q4) 
    (d2) [above, phitransitioncolor] edge node{ \failureterm{\smaller $\phi$} } (q5) 
    (q6) [above, phitransitioncolor] edge node{ \failureterm{\smaller $\phi$} } (d1) 
    (q7) [above, phitransitioncolor] edge node{ \failureterm{\smaller $\phi$} } (d2) ;
    
\end{tikzpicture}
\caption{A WFSA-$\phi$ fragment in which \cref{alg:new-backward-general} would perform a large number of updates over the $\phi$-transitions.}
\label{fig:shoelaces}
\end{figure}

The worst-case behavior is illustrated by \cref{fig:shoelaces}, where the only possible topological order is $1,2,3,4,5,\ldots$.  When \cref{line:general-topo} iterates over state 5 immediately after state 4, the aggregator must transition $4 \stackrel{\restorestep}{\longrightarrow} 2 \stackrel{\restorestep}{\longrightarrow} 1 \stackrel{\updatestep}{\longrightarrow} 3 \stackrel{\updatestep}{\longrightarrow} 5$.  Note that this involves 2 $\updatestep$s, as 2 is the height of state 5.

If the $a$ arcs were not present in \cref{fig:shoelaces}, however, then
$1,2,4,\ldots,3,5,\ldots$ would also be a topological order, which achieves the best-case behavior of visiting each state only once.  Indeed, many topological orders would be available---some more efficient than others.


\subsection{Topological Sorting Heuristics}\label{sec:toposort-heuristics}

It is desirable to choose a good topological order when one is available.  In particular, the ``best-case'' scenario above is achieved under this condition:

\begin{defin} 
  Let $\wfsa$ be an acyclic WFSA-$\phi$.  Given a reverse topological order of the states, we say that $q$ \defn{completely precedes} $q'$ if $q$ and all its failure-tree ancestors precede $q'$ and all its failure-tree ancestors.  We say that the order is \defn{compatible} with the failure trees of $\wfsa$ if whenever $q,q'$ are in the same failure tree\footnote{Remark: If we dropped the condition that $q,q'$ be in the same failure tree, then we would get a stronger compability criterion that would let all failure trees share a single aggregator.\looseness=-1} but have disjoint sets of ancestors, either $q$ completely precedes $q'$ or vice-versa.\looseness=-1
\end{defin}

To put this another way,  a compatible order of the WFSA states may jump back and forth among failure trees, as needed to achieve a topological ordering, but each failure tree's states will appear in some depth-first order starting at the tree's root, which ensures that each state is {\updatestep}ed just once.

In some backoff architectures such as variable-order conditional random fields \cite{vieira-cotterell-eisner:2016:EMNLP2016}, it is easy to find a compatible order.  In these WFSAs, each failure tree is associated with a position in a fixed input sentence.  Simply visit the failure trees from right to left, enumerating each one's states in depth-first order starting at the root.

For the general case, we have developed an topological sorting algorithm that will find a compatible order when one exists.  We begin with \citeauthor{kahn-1962}'s (\citeyear{kahn-1962}) agenda-based algorithm for finding a reverse topological order.  It places all states onto a very simple priority queue in which ``ready'' states are prioritized at the front of the queue.  The next state $q$ to enumerate is obtained by popping this queue, and then the parents of $q$ (that is, its immediate predecessors in the WFSA graph) decrement their counts of unenumerated children (i.e., immediate successors).  If a parent state's count reaches 0, then it becomes ready and moves to the front portion of the queue.  If the algorithm ever pops a non-ready state, then it throws an exception saying that the WFSA was cyclic.

Our approach is to modify Kahn's algorithm so as to break ties.  Once $q$'s children have been enumerated, Kahn's algorithm is allowed to enumerate $q$ at any time, but our modified version prefers to wait until it would be possible to enumerate $q$ and (eventually) its failure-tree ancestors with a single $\updatestep$ each.  Unfortunately, this test is expensive,\footnote{It must look ahead to determine whether any failure-tree ancestor of $q$ would have to wait for a state in some other part of the same failure tree.  As far as we know, this requires computing the transitive closure of the WFSA.} so using it would not actually speed up \cref{alg:new-backward-general}.  We therefore omit the details here.

In practice, we recommend using a greedy version of the above algorithm.  We do wait to enumerate $q$ until $q$ can be enumerated with a single $\updatestep$, but we no longer worry about its ancestors. This greedy heuristic is still guaranteed to find a compatible order if the WFSA has the special property that there are no paths between states in the same failure tree (other than $\phi$-paths).  Variable-order CRFs do have this property.  \Cref{fig:shoelaces} does not.

Specifically, we say that a not-yet-enumerated state $q \in \tree$ is \defn{cheap} if it is a $\phi$-parent of the current $q_\tree$ (that is, $\fallbackstate=q_\tree$), so that \cref{alg:new-backward-general} only has to call $\updatestep(\aggregator_{\tree},q)$ to update $q_\tree \gets q$.  Modify Kahn's algorithm to prioritize cheap ready states ahead of expensive ready states.\footnote{Each state (in particular $q_\tree$) maintains its own agenda of unenumerated $\phi$-parents, with the ready ones at the front.}  Modify \cref{alg:new-backward-general} to repeatedly descend at the end of the main loop until $q_\tree$ has at least one unenumerated $\phi$-parent,\footnote{We may descend all the way to the root of the failure tree without ever achieving this condition---but this implies that the failure tree has been fully enumerated, so there is no need to further update $\aggregator_\tree$ and $q_\tree$, as they will not be used again.} ensuring that there is a new cheap state in $\tree$.  The hope is that this cheap state will become ready while it is still cheap (indeed, it may already be ready).

\subsection{Copying Aggregators}

Long $\restorestep$--$\updatestep$ paths can trigger many updates to aggregator $\aggregator_\tree$.
Such paths can be shortened by splitting the failure tree into multiple smaller trees, each with its own aggregator.
When we $\updatestep$ a state $q$, we can choose to \emph{copy} the aggregator from $\fallbackstate$ and update only the copy, leaving the old aggregator at $\fallbackstate$.  While this incurs a one-time copying cost,
we can now split off the failure subtree rooted at $q$ into its own failure tree.  Enumerating states in this subtree will now never require visiting $q$'s descendants. The effect is to reduce $\maxfailuretreesize$ in the analysis of \cref{sec:new-backward-general-runtime}.  
\Cref{appendix:bisection-details} presents
\begin{itemize}[nosep]
\item a dynamic splitting heuristic that is sensitive to the actual toposort order (\cref{sec:toposort-heuristics})
\item an static splitting algorithm that uses dynamic programming to choose the \emph{optimal set} of split states to minimize a worst-case bound
\item runtime analysis of an idealized case to show how \cref{alg:new-backward-general} uses copying to gracefully degrade into \cref{alg:aggregation-backward} as the WFSA becomes denser
\end{itemize}

\subsection{The Ring Case}\label{sec:ring-aggregator}

In the case of a ring, it is possible to implement a faster aggregator. The aggregator still stores $N$ summands and their total, but no partial sums.  It can replace a summand in time $O(1)$ rather than $O(\log N)$, by subtracting off the old summand from the total and adding the new one.  This eliminates the $\log |\alphabet|$ factor from the runtimes in \cref{sec:new-backward-general-runtime}.

The resulting bound $\bigO ( |\transitions| + \avgoutfrac |\alphabet| |\states| \maxfailuretreesize)$ for \cref{alg:new-backward-general} is still worse than \cref{sec:ring-backward-runtime}'s bound of $\bigO{\left(|\transitions| + |\alphabet| |\states| \min(1,s\maxfailurepathsize) \right)}$ for \cref{alg:ring-backward}.  However, the former becomes better when a compatible order is known and the $\maxfailuretreesize$ can be dropped.

It is more instructive to compare the tighter bounds of $\bigO(|\transitions| + \sum_{q' \in \states} |\outsymbols(q')|\,\ancs(q'))$ for \cref{alg:new-backward-general} and $\bigO \left(|\transitions| + \sum_{q \in \states} |\hat{\outsymbols}(q)| \right)$ for \cref{alg:ring-backward}.  If a compatible order is known, $\ancs(q')$ can be dropped and the former is better.  If not, then either runtime could be better. The former effectively charges each state $q$ for all of the out-symbols at all of its $\phi$-descendants $q'$,\footnote{Since $\sum_{q' \in \states} |\outsymbols(q')|\,\ancs(q') = \sum_{q' \in \states} \sum_{q \preceq q'} |\outsymbols(q')| = \sum_{q \in \states} \sum_{q' \succeq q} |\outsymbols(q')|$.}
while the latter charges $q$ for all of the \emph{distinct} out-symbols at its $\phi$-\emph{ancestors} $q'$.
The reason for the difference: Both algorithms override a descendant's $a$-arc with an ancestor's $a$-arc, but to find these descendant-ancestor pairs, \cref{alg:new-backward-general} loops over $a$-arcs at the descendant (pushing subtrahends up from below) while \cref{alg:ring-backward} loops over $a$-arcs at the ancestor (pulling subtrahends up from above).  When different descendant-ancestor paths overlap, the former algorithm shares work between them if the $\updatestep$ order is good, while the latter shares work between them via memoization.



\section{Comparison of Algorithms}

This work proposed multiple algorithms for computing the pathsum of an acyclic WFSA-$\phi$. 
They are all alternatives to running the backward algorithm (\cref{alg:backward})---%
or its
simple improvement by aggregation (\cref{alg:aggregation-backward})---after explicitly expanding failure transitions (\cref{alg:failure-expansion}).
This section summarizes the improvements.

As mentioned in \cref{sec:naive-improvement}, we never change the way the local component of the backward values is computed. 
All algorithms we consider therefore retain the $\bigO{(|\transitions|)}$ complexity term from expanding the non-$\phi$ transitions.
What differs is the method for computing the failure term $\failureterm{\backwardweight(q, \alphabet \setminus \outsymbols(q))}$---the contribution of the paths starting at $q$ that take $q$'s failure transition.
\cref{tab:alg-cost-comparison} compares this term's runtime complexity for all the algorithms discussed.

\begin{table}
\renewcommand{\arraystretch}{1.2}
\centering
\begin{tabular}{lrl}
\toprule
\textbf{Alg.} & \textbf{$\bigO$-Cost} & \textbf{Use case}\\
\midrule
\cref{alg:backward}
&
$\failureterm{(\overline{\avgoutfrac} - s) |\states|} |\alphabet| |\states|$
&
never
\\ 
\cref{alg:aggregation-backward}
&
$\failureterm{(\overline{\avgoutfrac} - s)} |\alphabet| |\states|$
&
--
\\ 
\cref{alg:ring-backward}
&
$\failureterm{ \maxfailurepathsize \avgoutfrac} |\alphabet| |\states|$
&
$\avgoutfrac \ll \frac{\overline{\avgoutfrac} - s}{\maxfailurepathsize} $
\\ 
$\text{\cref{alg:new-backward-general}}^+$
&
$\failureterm{ \updatecomplexity\avgoutfrac } |\alphabet| |\states|$
&
$\avgoutfrac \ll \frac{\overline{\avgoutfrac} - s}{\updatecomplexity}$
\\ 
$\text{\cref{alg:new-backward-general}}^-$
&
$\failureterm{ \maxfailuretreesize \updatecomplexity \avgoutfrac} |\alphabet| |\states|$
&
$\avgoutfrac \ll \frac{\overline{\avgoutfrac} - s}{\maxfailuretreesize \updatecomplexity}$
\\ 
\cref{appendix:bisection-details}
&
$\failureterm{\sqrt{ \updatecomplexity\avgoutfrac} } |\alphabet| |\states| $
&
$\avgoutfrac \ll \frac{(\overline{\avgoutfrac} - s)^2}{\updatecomplexity}$
\\
\bottomrule
\end{tabular}
\caption{
Runtime of computing the \failureterm{failure term} by the different algorithms. The ``use case'' column indicates when an algorithm has better complexity than the baseline algorithm, \cref{alg:aggregation-backward}. $\updatecomplexity$ is the update complexity of the aggregator interface: $\log |\alphabet|$ in the general case (via a Fenwick tree) and $1$ in the ring case of \cref{sec:ring-aggregator} (via subtraction). $\text{\cref{alg:new-backward-general}}^+$ is the runtime for WFSA-$\phi$'s such as VoCRFs where a compatible state order is known, whereas $\text{\cref{alg:new-backward-general}}^-$ is the general worst-case runtime. 
}
\label{tab:alg-cost-comparison}
\end{table}

Maintaining perspective, the benefits of our more sophisticated pathsum algorithms over the basic \cref{alg:aggregation-backward} only make an actual impact if \cref{alg:aggregation-backward}'s failure component complexity $\bigO((\overline{\avgoutfrac} - \avgoutfrac) |\alphabet| |\states|)$ is dominant over the local component $\bigO(|\transitions|)$, where $|\transitions| \geq \avgoutfrac |\alphabet||\states|$.  In particular, reducing the failure component is only helpful if  $\overline{\avgoutfrac} \gg \avgoutfrac$, so that expanding failure transitions would make the graph denser.\looseness=-1



\section{Conclusion}

We presented two new algorithms for more efficiently computing the backward values and pathsum of a sparse acyclic semiring-weighted FSA with $\phi$-transitions, using the observation that a $\phi$-transition from $q$ to $\fallbackstate$ means that $\backwardweight(q)$ is a sparsely modified version of $\backwardweight(\fallbackstate)$.
We characterized when the new algorithms are asymptotically faster.

\section*{Limitations}
\label{sec:limitations}

This section addresses two main limitations of our work: the assumptions made on the structure of the WFSAs and the applicability of the proposed algorithms in real scenarios.

\paragraph{Acyclicity assumption.}
We only consider acyclic WFSAs.
While this covers interesting use cases such as CRFs, other commonly used instances of WFSAs also contain cycles, e.g., $n$-gram language models. 
Furthermore, all our novel algorithms actually assume that $\transitions \cup \transitions^\phi$ is acyclic, whereas failure expansion only requires that  the resulting $\failureexpansion$ is acyclic.
The former is a strictly stronger condition---see \cref{fig:failure-exp-acyclic-union-cyclic} below for an example WFSA-$\phi$ where $\transitions \cup \transitions^\phi$ is not acyclic, but $\failureexpansion$ is. \looseness=-1
\benji{If we allow cyclic phi transitions I think we can still failure expand - the runtime would just double. Each cycle of phi transitions would pick an arbitrary state as root, which would search out in one full cycle revolution its failure expanded edges set - the other states in the cycle would then just proceed as before in rev topo order starting from behind the root state.}

\paragraph{Applicability.}
As seen above, the runtime of \cref{alg:new-backward-general} depends on the size of failure trees, with complexity $\bigO ( |\transitions| + \avgoutfrac |\alphabet| |\states| \maxfailuretreesize  \log |\alphabet|)$.
In practice, failure trees may be large, or $s$ may be large, which could result in our algorithms performing worse than the na{\"i}ve approaches.\footnote{The hybrid approach proposed in \cref{appendix:bisection-details} aims at improving this shortcoming but requires additional knowledge of the WFSA's topology.}
To see this, consider higher-order CRFs with backoff, a useful formalism for sequence tagging in NLP \citep{vieira-cotterell-eisner:2016:EMNLP2016}, which can be encoded as WFSAs. 
They were the initial motivation for our proposed algorithms.
Although these backoff CRFs do admit a compatible topological order that allows us to avoid the $\maxfailuretreesize$ factor (\cref{sec:toposort-heuristics}), we inspect them as an example of how large $\maxfailuretreesize$ can be. \looseness=-1

An order-$n$ CRF tagging a sequence of length $\ell$ can be represented with a WFSA-$\phi$ in form of a lattice of $\ell$ layers. 
The layers include \emph{tag sequences} of length $\leq n$, meaning that, given a set of tags $\alphabet$, each layer contains states representing histories $h \in \{\epsilon\} \cup \alphabet \cup \cdots \cup \alphabet^n$.
This results in $\bigO{ \left ( |\alphabet|^n \right ) }$ states per layer.
Backoff transitions in such models encode transitions to lower-order histories (transitioning from a history of length $k$ to one of length $k-1$) whenever a transition to a history of the same order is not possible.
It is easy to see that each history of order $k$ could have up to $\alphabet$ incoming $\phi$-transitions, connecting it to a large failure tree, which is exponential in size w.r.t. $n$.

\section*{Ethics Statement}
We are not aware of any specific social risks created or exacerbated by this work.

\section*{Acknowledgements}
We would like to thank Alexandra Butoi for her valuable comments.\looseness=-1

\bibliographystyle{acl_natbib}
\bibliography{redux}

\appendix


\section{Algorithm Demonstrations} \label{appendix:algo-walkthrough}

Consider the WFSA-$\phi$ fragment in  \cref{fig:wfsa-fragment} and the version in \cref{fig:wfsa-fragment-expansion} that is produced by failure expansion (\cref{alg:failure-expansion}). 
This section demonstrates how different algorithms we discuss operate to compute the value $\backwardweight(q)$.
\begin{figure}
    \centering
    \begin{subfigure}[b]{\linewidth}
        \centering
        \begin{tikzpicture}[node distance = 12 mm]
        \footnotesize
        \node[state] (q) [] { $q$ }; 
        \node[state] (qphi) [right = 10 mm of q] { $\fallbackstate$ }; 
        \node[state] (q1) [above left = 10mm of q] { $q_1$ }; 
        \node[state] (q2) [below left = 10mm of q] { $q_2$ }; 
        \node[state] (q3) [above right = 10mm of qphi] { $q_3$ }; 
        \node[state] (q4) [below right = 10mm of qphi] { $q_4$ }; 
        \draw[-{Latex[length=3mm]}] 
        (q) edge[above, swap, phitransitioncolor] node{ $\phi / \semione$ } (qphi)
        (q) edge[above right, swap, localtermcolor] node{ $a / w_{1}$ } (q1)
        (q) edge[below right, bend left, swap, localtermcolor] node{ $b / w_{2b}$ } (q2)
        (q) edge[above left, bend right, swap, localtermcolor] node{ $a / w_{2a}$ } (q2)
        (qphi) edge[below right, swap, failuretermcolor] node{ $c / w_{3}$ } (q3)
        (qphi) edge[below left, swap, failuretermcolor] node{ $b / w_{4}$ } (q4);
        \end{tikzpicture}
        \caption{A fragment inside of a WFSA-$\phi$ graph.}
        \label{fig:wfsa-fragment}
    \end{subfigure}
    \par\bigskip\medskip
    \begin{subfigure}[b]{\linewidth}
        \centering
        \begin{tikzpicture}[node distance = 12 mm]
        \footnotesize
        \node[state] (q) [] { $q$ }; 
        \node[state] (qphi) [right = 10 mm of q] { $\fallbackstate$ }; 
        \node[state] (q1) [above left = 10mm of q] { $q_1$ }; 
        \node[state] (q2) [below left = 10mm of q] { $q_2$ }; 
        \node[state] (q3) [above right = 10mm of qphi] { $q_3$ }; 
        \node[state] (q4) [below right = 10mm of qphi] { $q_4$ }; 
        \draw[-{Latex[length=3mm]}] 
        (q) edge[above left, swap, failuretermcolor] node{ $c / w_3$ } (q3)
        (q) edge[above right, swap, localtermcolor] node{ $a / w_{1}$ } (q1)
        (q) edge[below right, bend left, swap, localtermcolor] node{ $b / w_{2b}$ } (q2)
        (q) edge[above left, bend right, swap, localtermcolor] node{ $a / w_{2a}$ } (q2)
        (qphi) edge[below right, swap, failuretermcolor] node{ $c / w_{3}$ } (q3)
        (qphi) edge[below left, swap, failuretermcolor] node{ $b / w_{4}$ } (q4);
        \end{tikzpicture}
        \caption{Failure-expanded version of the fragment from \cref{fig:wfsa-fragment}.}
        \label{fig:wfsa-fragment-expansion}
    \end{subfigure}
    \vspace{6pt}
    \caption{WFSA-$\phi$ and WFSA examples discussed in \cref{appendix:algo-walkthrough}.}
\end{figure}

The normal backward algorithm (\cref{alg:backward}) on the failure-expanded version would compute $\backwardweight(q)$ as 
\begin{align*}
    \backwardweight(q) &\gets \localterm{w_1 \otimes \backwardweight(q_1) \oplus w_{2a} \otimes \backwardweight(q_2)} \\ 
    & \qquad \localterm{\mbox{}\oplus w_{2b} \otimes \backwardweight(q_2)} \failureterm{\mbox{}\oplus w_3 \otimes \backwardweight(q_3)}.
\end{align*}

The version that memoizes out-symbol sums (\cref{alg:aggregation-backward}) would compute $\backwardweight(q)$ as 
\begin{align*}
    \backwardweight(q) &\gets \localterm{w_1 \otimes \backwardweight(q_1) \oplus w_{2a} \otimes \backwardweight(q_2)} \\ 
    & \qquad \localterm{\mbox{} \oplus w_{2b} \otimes \backwardweight(q_2)} \failureterm{\mbox{} \oplus \backwardweight(\fallbackstate, c)}.
\end{align*}
\Cref{alg:aggregation-backward} is equivalent to copying the entire $\backwardweight(\fallbackstate, a)$ memo table from $\fallbackstate$, modifying the values for $a \in \outsymbols(q)$, and summing.
That is, the dictionary $\{c\mapsto w_3 \otimes \backwardweight(q_3), b\mapsto w_4 \otimes \backwardweight(q_4)\}$ would be passed back from $\fallbackstate$ to $q$ and updated there to $\{c\mapsto w_3 \otimes \backwardweight(q_3), b\mapsto w_{2a} \otimes \backwardweight(q_2) \oplus w_{2b} \otimes \backwardweight(q_2), a\mapsto w_1 \otimes \backwardweight(q_1)\}$, and $\backwardweight(q,\alphabet)$ would be found by summing the values in this dictionary.

The subtraction-based algorithm (\cref{alg:ring-backward}) would compute $\backwardweight(q)$ as 

\noindent
\begin{align*}
    \backwardweight(q) &\gets \localterm{w_1 \otimes \backwardweight(q_1) \oplus w_{2a} \otimes \backwardweight(q_2)} \\ 
    & \qquad \localterm{\mbox{}\oplus w_{2b} \otimes \backwardweight(q_2)} \failureterm{\mbox{}\oplus \backwardweight(\fallbackstate) \ominus \backwardweight(\fallbackstate, \{a, b\})} \\
    &=\localterm{w_1 \otimes \backwardweight(q_1) \oplus w_{2a} \otimes \backwardweight(q_2)} \\ 
    & \qquad \localterm{\mbox{}\oplus w_{2b} \otimes \backwardweight(q_2)} \failureterm{\mbox{}\oplus \backwardweight(\fallbackstate) \ominus \backwardweight(\fallbackstate, \{b\})}.
\end{align*}

Lastly, \cref{alg:new-backward-general} would initialize an aggregator $\aggregator$ at the failure tree root $\fallbackstate$ as $\{c\mapsto w_3 \otimes \backwardweight(q_3), b\mapsto w_4 \otimes \backwardweight(q_4)\}$, and pass the aggregator back to $q$.
There, $\aggregator$ would be updated via $\aggregator.\myfunc{set}(a,\ldots)$ and $\aggregator.\myfunc{set}(b,\ldots)$ to $\{c\mapsto w_3 \otimes \backwardweight(q_3), b\mapsto w_{2a} \otimes \backwardweight(q_2) \oplus w_{2b} \otimes \backwardweight(q_2), a\mapsto w_1 \otimes \backwardweight(q_1)\}$, causing $\aggregator.\myfunc{value}()$ to change.  Then, $\backwardweight(q)$ would be computed as $\backwardweight(q) = \aggregator.\myfunc{value}()$.
Compare this to \cref{alg:aggregation-backward}, which had to explicitly sum up all the values in this dictionary to compute $\backwardweight(q)$, since it did not use an aggregator data structure (\cref{appendix:fenwick}) to maintain partial sums over subsets of these values.

\section{Number of Transitions Added by Failure Expansion}
\label{appendix:failure-expansion-num-added-transitions}
We show \cref{sec:failure-expansion}'s claim that the number of transitions added by failure expansion (\cref{alg:failure-expansion}) is $(\overline{\avgoutfrac} - \avgoutfrac) |\alphabet| |\states|$ when the input WFSA-$\phi$ is deterministic. 

In the deterministic case, each out-symbol at a state labels exactly one outgoing transition.  Hence the number of added transitions for a given state $q$ equals the number of added out-symbols, $|\overline{\outsymbols}(q) \setminus \outsymbols(q)| = |\overline{\outsymbols}(q)| - |\outsymbols(q)|$, where $\overline{\outsymbols}(q) \supseteq \outsymbols(q)$. Summing over all $q \in \states$, and using \cref{def:avgoutfrac}, we get a total number of added transitions of 
\begin{align*}
    \sum_{q \in \states} \overline{\outsymbols}(q) - \sum_{q \in \states} \outsymbols(q) &= \overline{\avgoutfrac} |\alphabet| |\states| - \avgoutfrac |\alphabet| |\states|
    \\
    &= (\overline{\avgoutfrac} - \avgoutfrac) |\alphabet| |\states|.
\end{align*}

In the general case where the input WFSA-$\phi$ may be non-deterministic, each added out-symbol may label anywhere from 1 to $|\states|$ added transitions.  Thus the total number of added transitions is between $(\overline{\avgoutfrac} - \avgoutfrac) |\alphabet| |\states|$ and  $(\overline{\avgoutfrac} - \avgoutfrac) |\alphabet| |\states|^2$.

\section{Aggregator Implementation} \label{appendix:fenwick}
A Fenwick tree \citep{fenwick1994new} is a data structure that stores a
sequence $v_1, \ldots, v_N$ and can efficiently return any prefix sum
of the form $\bigoplus_{n = 1}^{N'} v_n$ for $N' \in [0,N]$, as well
as allowing the individual elements $v_n$ to be updated.  Each prefix-sum query
or element update takes $\bigO(\log N)$ time.  

Our aggregator interface in \cref{sec:aggregator} is simpler.  It only
queries the full sum $\bigoplus_{n = 1}^{N} v_n$ (the case $N'=N$).
Thus, the order of the elements is not considered by this interface.  \Cref{sec:ring-aggregator} noted that in the
special case where subtraction is available (and numerically stable),
an aggregator can be implemented even more efficiently \emph{without}
a Fenwick tree, since then it is easy to update the sum in
constant time when updating any element.  However, subtraction is not
guaranteed to be available for arbitrary $\oplus$ operations (e.g., $\oplus=\max$).

A Fenwick tree stores the elements $v_n$ at the leaves of a balanced binary tree.
Each internal (non-leaf) node stores the $\oplus$-sum of the values stored at its children.  As a result, thanks to the associativity of $\oplus$, the root of the tree contains the full sum $\bigoplus_{n = 1}^N v_n$, which can be looked up in $\bigO(1)$ time.
An example of a Fenwick tree (in the real semiring) is presented in \cref{fig:fenwick-tree}.  Note that we draw the root of a Fenwick tree at the top and consider it to be the ancestor of all other nodes, whereas failure trees had the root as the descendant of all other states. 
\begin{figure}
    \centering
    \scalebox{0.7}{\begin{tikzpicture}
    \node[circle,draw](z){$6$}
        child{
            node[circle,draw]{$3$} 
                child{
                    node[circle,draw] {$1$}
                } 
                child{
                    node[circle,draw] {$2$}
                } 
        }
        child{
            node[circle,draw]{$3$} 
        };
    \end{tikzpicture}}
    \caption{A Fenwick tree computing $1 + 2 + 3 = 3 + 3 = 6$.}
    \label{fig:fenwick-tree}
\end{figure}

Initial creation of the Fenwick tree takes only $\bigO(N)$ total time by visiting all nodes in bottom-up order and setting each non-leaf node to the $\oplus$-sum of its children.  When a leaf $v_n$ is updated, just its ancestors are recomputed, again in bottom-up order.  As there are about $\log N$ ancestors, this update takes $\bigO(\log N)$ total time.


Our aggregator is a Fenwick tree that stores $N=|\alphabet|$ elements,
where $v_n$ is the value associated with the $n$\textsuperscript{th}
element of $\alphabet$.  (That is, we identify the possible keys
$a \in \alphabet$ with the integers $[1,N]$.)  Initially,
$v_n=\semizero$, but may be changed by \myfunc{set}.  Each call to
\myfunc{set} takes $\bigO(\log |\alphabet|)$ time; this factor appears
in our runtime analysis.  To achieve our runtime bounds for sparse
WFSAs, we must take care not to spend $\bigO(|\alphabet|)$ time
initializing all of the leaves and internal nodes to $\semizero$ every
time we create an aggregator.  Array initialization overhead can
always be avoided, using a method from computer 
science folklore \cite[exercise 2.12]{aho-hopcroft-ullman-1974}.

Alternatively, we can store values in the Fenwick tree only for those keys for which values have been set.  Under this design, the operation 
$\myfunc{set}(a: \alphabet, v: \semiringset)$ must update $v_n \gets v$ where $n$ is the integer index associated with key $a$.  To find $n$, the aggregator maintains a hash table that maps keys to consecutive integers.  We assume $\bigO(1)$-time hash operations.  The first key that is set is mapped to 1, the second is mapped to 2, etc.  When a key $a$ is set for the first time---that is, when it is not found in the hash table---$N$ is incremented, the mapping $a \mapsto N$ is added to the hash table, and $v_N = v$ is appended to the Fenwick sequence.  The hash table is also consulted by the \myfunc{get} operation.

In our application, for an aggregator that represents state $q$, the keys that have been set are $\extendedoutsymbols(q)$.  The design in the previous paragraph therefore reduces $N$ from $|\alphabet|$ to $N=|\extendedoutsymbols(q)|$.  As a result, the factor $\bigO(\log |\alphabet|)$ in our analysis could actually be reduced to $\bigO(\log \max_{q \in \states} |\extendedoutsymbols(q)|)$.

Note that to obtain this runtime reduction, the \myfunc{undo} method must properly undo the changes not only to the Fenwick tree but to the integerizing hash table (see \cref{fn:undo}).  If a call to \myfunc{set} in $\updatestep$ incremented $N$ and added $a \mapsto N$, then the call to \myfunc{undo} in $\restorestep$ must remove $a \mapsto N$ and decrement $N$, thereby keeping $N$ small as desired.


\section{Weighted $\phi$-Transitions}\label{appendix:failure-weight}
Throughout the main paper, we assumed that all $\phi$-transitions have a weight of $\semione$.
This simplifying assumption is typically violated by backoff models \cite[e.g.,][]{allauzen-mohri-roark:2003:ACL}.  Fortunately, it can be removed with relatively small changes to our equations, algorithms and data structures.  

Most simply, a weighted failure transition $\edge{q}{\phi}{w^\phi}{\fallbackstate}$ could be simulated by a path $\edge{q}{\phi}{\semione}{\mbox{}}\edge{q^\varepsilon}{\varepsilon}{w^\phi}{\fallbackstate}$ where $q^\varepsilon$ is a newly introduced intermediate state with only an $\varepsilon$-transition.  We would then have to eliminate the $\varepsilon$-transition as mentioned in  \cref{sec:preliminaries}.  In this case, this simply means replacing $\edge{q^\varepsilon}{\varepsilon}{w^\phi}{\fallbackstate}$ in $\transitions$ with transitions $\{\edge{q^\varepsilon}{a}{w^\phi \otimes w}{q'}: \edge{\fallbackstate}{a}{w}{q'} \in \transitions\}$.  However, this may be expensive when the original fallback state $\fallbackstate$ has many outgoing transitions, which is typical in a backoff setting.  Copying all of those transitions to a parent as in \cref{alg:failure-expansion} (failure expansion) is exactly what the new methods in this paper are designed to avoid.  We therefore give direct modifications to our constructions.

Suppose the failure transition for state $q$ has weight $w^\phi$---that is, $\transitions$ contains $\edge{q}{\phi}{w^\phi}{\fallbackstate}$---where perhaps $w^\phi \neq \semione$.
Then the second case of \cref{eq:backwardweight-phi-state-symbol-decomposition} should be modified to set 
\begin{align*}
\backwardweight(q, a) &= w^\phi \otimes \backwardweight(\fallbackstate, a)
\end{align*}
for any $a \notin \outsymbols(q)$. Similarly, $w^\phi$ should be incorporated into 
\cref{eq:backward-sigma-sum}, which becomes
\begin{align*}
    \failureterm{\backwardweight(q, \alphabet \setminus \outsymbols(q))}
    &= w^\phi \otimes \monsteroplus[r]{b \in \alphabet \setminus \outsymbols(q)} \backwardweight(\fallbackstate, b)
\end{align*}
Finally, the subtraction expression in the right-hand side of \cref{eq:backward-minus} must be left-multiplied by $w^\phi$.  
 
In \cref{alg:failure-expansion}, which constructs the failure-expanded edge set, the update at state $q$ becomes
\begin{align*}
\failureexpansion &\gets \failureexpansion \cup \{ \edge{q}{a}{w^\phi \otimes w}{q'} \mid \edge{\fallbackstate}{a}{w}{q'} \in \failureexpansion, a \notin \outsymbols(q) \}
\end{align*}


\Cref{alg:aggregation-backward,alg:ring-backward} undergo straightforward modifications 
based on the modified \cref{eq:backwardweight-phi-state-symbol-decomposition,eq:backward-sigma-sum,eq:backward-minus}.
When  $\backwardweight(\fallbackstate, b)$ is copied backwards over a $\phi$-transition $\edge{q}{\phi}{w^\phi}{\fallbackstate}$, it must be left-multiplied by $w^\phi$ to yield $\backwardweight(q, b)$.  This affects \cref{alg:aggregation-backward} \crefrange{line:aggregation-backward-failure-copy}{line:aggregation-backward-failure-sum} and \cref{alg:ring-backward} \cref{line:ring-recurse}, as well as the purple terms in \cref{alg:ring-backward} \cref{line:ring-backward-replace}.
These modifications do not affect the asymptotic runtime complexity.

\cref{alg:new-backward-general} requires more modification.  We must extend our aggregator class (\cref{sec:aggregator}) with a new method that left-multiplies all elements by a constant:\footnote{This extension can more generally support the case where the multipliers $m$ fall in a monoid of scalars $(\mathbb{M},\otimes,\semione)$ that acts on a monoid  $(\semiringset,\oplus,\semizero)$ of values $v$ (perhaps vectors).  However, for our WFSA-$\phi$ application, both the scalars and the values are weights from the same semiring: $m$ will always be a failure weight $w^\phi \in \semiringset$.}
\begin{algorithm}[h]
\begin{algorithmic}[1]
\State \textbf{class} \aggregatortype{}(): \InlineComment{We use $\aggregator$ to refer to an aggregator instance}

$\qquad\vdots$
\alglinenumRestore{aggregator}
\State \quad\textbf{def} \myfunc{mult}($m$: $\semiringset$) \InlineComment{$\forall a \in \alphabet$, updates $\aggregator(a) \gets m \otimes \aggregator(a)$}
\end{algorithmic}
\end{algorithm}

In \cref{alg:update-restore}, $\updatestep(\aggregator,q)$ should begin by calling $\myfunc{mult}(w^\phi)$ where $w^\phi$ is the weight of the failure arc from $q$.  Consequently, $\restorestep(\aggregator,q)$ should be modified to $\myfunc{undo}$ one more update than before.

How to implement the $\myfunc{mult}$ method efficiently?  

\paragraph{With both subtraction and division} The subtraction-based aggregator (\cref{sec:ring-aggregator}) can be modified to still support all operations in $\bigO(1)$ time, provided that the ring $\semiringset$ is actually a divsion ring (noncommutative field), i.e., it supports division by non-$\semizero$ multipliers.  The aggregator maintains an overall multiplier $M$, initially $\semione$, and the call $\myfunc{mult}(m)$ replaces $M$ with $m \otimes M$; thus, $M$ is a product of the $I$ multipliers applied far, $m_I \otimes \cdots \otimes m_1$.  As in \cref{appendix:fenwick}, we identify each key with an integer index $n$.  If $a$ has index $n$, then $\myfunc{set}(a,v)$ stores $M^{-1} \otimes v$ into $v_n$.\footnote{In the special case where $M=\semizero$, so that $M^{-1}$ does not exist, an arbitrary value may be stored, since it will just be multiplied by $\semizero$ later.}
Later $\myfunc{get}(a)$ can return $M \otimes v_n$; since $M$ has been updated in the meantime, this yields the originally set value $v$ left-multiplied by all \emph{subsequent} multipliers, since $M$ has been updated.  The aggregator also maintains the total $\bigoplus_{n=1}^N v_n$ as $v_n$ values are set or replaced (using subtraction), and the $\myfunc{value}$ method returns $M \otimes \bigoplus_{n=1}^N v_n$.

\paragraph{With subtraction only} If $\semiringset$ does not support division, then the subtraction-based aggregator can be rescued as follows.  The aggregator maintains the number $I$ of multipliers applied so far, as well as their product $M$ as before.  The function $\myfunc{set}(a,v)$ now stores $v$ into $v_n$ and $I$ into $i_n$, and later $\myfunc{get}(a)$ returns $M_{i_n} \otimes v_n$, where in general $M_i$ is defined to be the product of multipliers \emph{subsequent} to $m_i$, that is, $M_i = M_I \otimes \cdots \otimes m_{i+1}$.  The aggregator maintains the current total $S$ that should be returned by $\myfunc{lookup}$; the $\myfunc{mult}(m)$ method left-multiplies this total by $m$, while the method $\myfunc{set}(a,v)$ modifies this total by adding $v \ominus \myfunc{get}(a)$ before it updates $(v_n,i_n)$.\footnote{Although \myfunc{get} is used internally within this implementation of \myfunc{set}, that does not contradict the comment later in this section that \myfunc{get} is never called directly by \crefrange{alg:update-restore}{alg:new-backward-general}.}  The difficulty is now in obtaining the partial products $M_i$ without division.  This can be done by maintaining $m_1,\ldots,m_I$ in a Fenwick tree.\footnote{A Fenwick tree supports suffix sums (or indeed, sums over any contiguous subsequence of elements) as efficiently as prefix sums.  In our case, the elements in the Fenwick tree are multipliers, so the ``sum'' operation to be used is $\otimes$, not $\oplus$.\looseness=-1}  This means that $\myfunc{mult}$ and $\myfunc{get}$ now take time $\bigO(\log I)$ rather than $\bigO(1)$.  The effect on \cref{sec:ring-aggregator}'s ring-based version of \cref{alg:new-backward-general} is to add $\sum_{q'\in\states} \ancs(q') \log \maxfailurepathsize \leq |\states| \maxfailuretreesize \log \maxfailurepathsize$ to the asymptotic runtime expression.  This is the same cost as if every state had $\log \maxfailurepathsize$ additional outgoing symbols.  $\maxfailurepathsize$ is usually very small.

\paragraph{Without subtraction}
For this case, we stored the summands in a Fenwick tree  (\cref{appendix:fenwick}).  Fortunately, it is possible to extend that data structure to support \myfunc{mult} in time $\bigO(N)$, where $N$ is the number of elements, without affecting the asymptotic runtime of \myfunc{set}, \myfunc{value}, or \myfunc{undo}.  The asymptotic runtimes of \crefrange{alg:update-restore}{alg:new-backward-general} will remain unchanged.


In our modified Fenwick tree, the $N$ leaves store \emph{unscaled} values $u_1,\ldots,u_N\in\semiringset$.  Each node $j$ (leaf or internal node) stores a multiplier $m_j$ that will be lazily applied to all of the leaves that are descendants of $j$.  Thus, the scaled value $v_n$ is found as the product $m_r \otimes m_{j_1} \otimes m_{j_2} \otimes \cdots \otimes m_n \otimes u_n$, where $r,j_1,\ldots n$ is the path from the root $r$ to the leaf $n$.  Thus, the leaves  store the elements $v_n$ directly (as they would in an ordinary Fenwick tree) only in the special case where all the multipliers are $\semione$. In general $v_n$ must be computed on demand.  The runtime of \myfunc{get} is now $\bigO(\log N)$ rather than $\bigO(1)$, but \crefrange{alg:update-restore}{alg:new-backward-general} never actually use the \myfunc{get} method.

The new call $\myfunc{mult}(m)$ simply replaces 
$m_r \gets m \otimes m_r$, which affects all $v_n$ in $\bigO(1)$ total time.

To support fast computation of the total value $\oplus_{n=1}^N v_N$, we also store partial sums at the nodes, as before.  Thus, each node $j$ stores a pair $(m_j,u_j)$.  The \emph{scaled value} of node $j$ is defined to be $m_j \otimes u_j$.  When $j$ is an internal node, we ensure as an invariant that $u_j$ is the sum of the scaled values of $j$'s children, updating it whenever $j$'s children change.  
The $\myfunc{value}$ method simply returns the scaled value of the root in $\bigO(1)$ time.

The interesting modification is to the \myfunc{set} method.  To set $v_n$ to $v$, leaf $n$ is modified to set $(m_n,u_n) \gets (\semione,v)$---but also, all of $n$'s ancestors $j$ must be modified to have multipliers $m_j=\semione$, so that $v_n = \semione \otimes \cdots \otimes \semione \otimes v = v$ as desired. Before being set to $\semione$, each old multiplier $m_j$ is ``pushed down'' to its children so that it still affects all leaves of $j$.  The method descends from the root $r$ to leaf $n$: it pushes the $m_j$ values out of the way on the way down, updates the leaf at the bottom, and restores the invariant by recomputing the $u_j$ values on the way back up as it returns. 

\begin{algorithmic}[1]
\Func{\myfunc{set}($a$: $\alphabet$, $v$: $\semiringset$)} 
\State $n \gets \mbox{}$leaf that stores value of key $a$
\State \myfunc{set\_desc}($n,v,r$) \InlineComment{$r$ is the root of the Fenwick tree}
\EndFunc
\Func{\myfunc{set\_desc}($n$: leaf, $v$: $\semiringset$, $j$: node)}
\LineComment{$j$ is an ancestor of $n$; $j$'s own proper ancestors have multiplier $\semione$; so will $j$ upon return}
\If{$j$ is a leaf}
$(m_j,u_j) \gets (\semione,v)$\InlineComment{Since $j=n$}
\Else
\alglinenumNew{fenwickmult}
\alglinenumSave{fenwickmult}
\For{$k \in \myfunc{children}(j)$} 
    $m_k\gets m_j \otimes m_k$
\EndFor
\State $m_j \gets \semione$\InlineComment{$m_j$ has been pushed down}
\State \myfunc{set\_desc}($n$, $v$, child of $j$ that is anc.\@ of $n$)
\State $u_j \gets \bigoplus_{k \in \myfunc{children}(j)} m_k \otimes u_k$ \InlineComment{Restore invariant at $j$}
\EndIf
\EndFunc
\end{algorithmic}
The $\mathbf{else}$ clause in \myfunc{set\_desc} can be rephrased (less readably) to avoid looping twice over $\myfunc{children}(j)$:
\begin{algorithmic}[1]
\alglinenumRestore{fenwickmult}\Indent\Indent
\State $k \gets \mbox{}$the child of $j$ that is an ancestor of $n$
\State $m_k\gets m_j \otimes m_k$ \InlineComment{push $m_j$ down to $k$}
\State \myfunc{set\_desc}($n,v,k$)
\State $u_j \gets u_k$ \InlineComment{$\mbox{}=m_k \otimes u_k$, since now $m_k=\semione$}
\For{$k' \in \myfunc{siblings}(k)$}\InlineComment{in a binary tree, there will be $\leq 1$}
    \State $m_{k'}\gets m_j \otimes m_{k'}$ \InlineComment{push $m_j$ down to $k'$}
    \State $u_j \opluseq m_{k'} \otimes u_{k'}$
\EndFor
\State $m_j \gets \semione$  \InlineComment{$m_j$ has been pushed down and invariant restored at $j$}
\EndIndent\EndIndent
\end{algorithmic}

\section{Tree Splitting Details}\label{appendix:bisection-details}
\cref{alg:new-backward-general} is applicable to any acyclic semiring-weighted WFSA-$\phi$.
However, updating an \aggregatortype{} as it travels within a failure tree incurs an additional worst-case multiplicative runtime factor of $\maxfailuretreesize$, the size of the biggest failure tree. 
This section outlines an improvement by lessening this impact.
We do so by \emph{splitting} large failure trees into multiple smaller ones.\looseness=-1

\Cref{alg:new-backward-general} destructively updates 
an aggregator $\aggregator$ when $\updatestep$ing a state $q$ from $\fallbackstate$.  This takes time $\bigO(|\outsymbols(q)| \log |\alphabet|)$.  In contrast, \cref{alg:aggregation-backward} can be thought of as non-destructively \emph{copying} $\aggregator$ to $q$ from $\fallbackstate$, which means the work can be saved and does not have to be redone if $q$ is re-$\updatestep$ed later.

This inspires us to hybridize \cref{alg:new-backward-general} as follows:
$\updatestep(\aggregator,q)$ in \cref{alg:update-restore} may optionally copy-and-update $\aggregator$ rather than just updating it.  
Copying effectively cuts the transition $\edge{q}{\phi}{w}{\fallbackstate}$, making the sub-tree rooted at $q$ a new independent failure tree with its own \aggregatortype{} instance.
Copy-and-update does incur a one-time cost of $\bigO{(|\alphabet|)}$,\footnote{\label{fn:copy-and-update}When we copy-and-update, each modification takes only $\bigO(1)$ time, not $\bigO(\log |\alphabet|)$.  The strategy is to copy all $\bigO(|\alphabet|)$ elements from $\aggregator$ of the Fenwick tree, update some of them, and only then build a new Fenwick tree from the updated elements, which takes only $\bigO(|\alphabet|)$ time: see \cref{appendix:fenwick}.}
but now \cref{alg:new-backward-general} \cref{line:choose-tree} will select a smaller failure tree.  

However, at what states (if any) should we split each failure tree?  The optimal set of splits depends on the topological order (\cref{sec:toposort-heuristics}) used by \cref{alg:new-backward-general}.

\paragraph{Dynamic splitting heuristics}
A simple greedy heuristic would be to split at $q$ upon any call $\updatestep(q)$ where copy-and-update is estimated to be cheaper than destructive updating, based on the current size of the aggregator and the number of required updates $|\outsymbols(q)|$.  
However, this does not consider the future benefit of having smaller failure trees, and it does not adapt to the topological order.

A more sophisticated dynamic heuristic is for $\updatestep(q)$ to split at $q$ if not doing so would cause the \emph{total} time spent so far on all $\updatestep(q)$ calls\footnote{Remark: For a given $q$, all such calls do exactly the same work and should take the same amount of time, regardless of other splits.} to exceed the time that it would take to copy-and-update at $q$.  (Put another way, it does so if it now realizes in retrospect that it would have been better for the very first $\updatestep(q)$ call to have invested in copy-and-update.)  This ensures that our enhanced \cref{alg:new-backward-general} will take at most twice as long as \cref{alg:aggregation-backward}, which always does copy-and-update.  It eventually splits any state that is $\updatestep$ed often enough by the chosen topological order, especially if that state is expensive to $\updatestep$.  On the other hand, if the chosen topological order is compatible so that every state is $\updatestep$ed only once, it will still achieve or outperform the best-case behavior of the original \cref{alg:new-backward-general}.

\paragraph{Static splitting algorithms}
We may also consider static methods, which do not adapt to the topological order that is actually used, but optimize to mitigate the worst case.
In the runtime analysis of \cref{sec:new-backward-general-runtime}, the failure tree $\tree$ contributes
$\bigO(f(\tree) \log |\alphabet|)$
to the failure term in the worst-case runtime of \cref{alg:new-backward-general},\footnote{Versus 
$\bigO(\sum_{q \in \tree} |\overline{\outsymbols}(q)| - |\outsymbols(q)|)$
in the case of \cref{alg:aggregation-backward}.}  
where $f(\tree) \defeq \sum_{q \in \tree} |\outsymbols(q)|\,\ancs(q)$.\footnote{\label{fn:skiproot-followup}As \cref{fn:skiproot} notes, this bound may be improved by defining $f$ to sum over only the \emph{non-root} states $q \in \tree$.}
%
%
We may seek a split that is optimal with respect to this runtime bound.  Let $q_1$ be the root of $\tree$.  Suppose we choose to copy-and-update the aggregator when we first visit each of $q_2, \ldots, q_K \in \tree$, essentially cutting off each state $q_k$ from its fallback state $q_k^\phi$.  (Here all of the $q_k$ are to be distinct.)  This splits $\tree$ into trees $\tree_1, \ldots, \tree_K$, where each $\tree_k$ is rooted at $q_k$. 
Then the contribution of these $K$ trees to the asymptotic runtime upper bound is proportional to $(K-1)|\alphabet| + 
\sum_{k=1}^K f(\tree_k) \log |\alphabet|$, 
where the first term covers the cost of the $K-1$ copy-and-update operations, and where the factor $\ancs(q)$ in the definition of $f(\tree_k)$ considers only the ancestors of $q$ within $\tree_k$.  Our goal is to choose $K \geq 1$ and $q_2, \ldots, q_K$ to minimize this expression.


We first remark that requiring $K \leq 2$ makes it easy to solve the problem in time $\bigO(|\tree|)$ time, assuming that we already know $|\Sigma(q)|$ for each $q \in \tree$. 
Define $D_q = \sum_{q' \succ q} |\outsymbols(q')|$, the total number of out-symbols at proper descendants of $q$.  The improvement $f(T) - (f(T_1)+f(T_2))$ from splitting $\tree$ at $q_2$ is simply $D_{q_2} \ancs(q_2)$.  Intuitively, there are $D_{q_2}$ out-symbols that can no longer be encountered when $\updatesteps$ is called on any of the $\ancs(q_2)$ states in 
$\tree_2$.\footnote{When $f$ is defined using the tighter bound of \cref{fn:skiproot,fn:skiproot-followup}, redefine $D_q = \sum_{q_1 \succ q' \succeq q} |\outsymbols(q')|$, so that $D_{q_2}$ will now include $q_2$ but exclude $q_1$.}  This yields an improvement of $-|\alphabet|+D_{q_2} \ancs(q_2)\log |\alphabet|$ in the runtime bound.  A simple recursion from the root $q_1$ is enough to find $D_q$ and $\ancs(q)$ at every state $q$, and thus find the state $q \neq q_1$ that achieves the best improvement in the runtime bound when chosen as $q_2$.  If no choice achieves a positive improvement, then we do not split the tree and leave $K=1$.

We now present an exact algorithm for the full problem, with no bound on $K$.  Roughly speaking, after we split at a state $q'$ (making it the root of its own failure tree), we will also consider splitting again at its ancestors $q$, but we do not make these decisions greedily---we use dynamic programming.  The main observation is that if $q$ is currently in a failure tree with root $q' \succ q$ (where either $q'=q_1$ or we previously split at $q'$), then splitting at $q$ will give a further improvement of $-|\alphabet| + (D_{q} - D_{q'}) \ancs(q) \log |\alphabet|$.  Denote this quantity by $\Delta_{q|q'}$.  We now wish to find the set of states $S = \{q_2,\ldots,q_K\} \subseteq \tree \setminus \{q_1\}$ that maximizes $\sum_{k=2}^K \Delta_{q_k \mid q'_k}$, where $q'_k$ is the highest state in $\{q_1,\ldots,q_K\}$ that is a proper descendant of $q_k$ (that is, $q'_k \succ q_k)$.  This sum is the total improvement obtained by splitting at all of $\{q_2,\ldots,q_K\}$, since it is the total that would be obtained by splitting them successively in any reverse topological order.

For each state $q \in \tree$ and each $q' \succ q$, define
\begin{align}
\bar{\Delta}_{q|q'} &= \mathrlap{\max(\check{\Delta}_{q|q'},\hat{\Delta}_{q|q'})} \label{eq:deltabar}\\
\check{\Delta}_{q|q'} &=  \left(\textstyle\sum_p \bar{\Delta}_{p|q}\right) + \Delta_{q|q'} \label{eq:deltacheck} \\
\hat{\Delta}_{q|q'} &= \left(\textstyle\sum_p \bar{\Delta}_{p|q'}\right) + 0 \label{eq:deltahat}
\end{align}
where $p$ in the summations ranges over the parents of $q$ (if any) in the failure tree.  Here $\bar{\Delta}_{q|q'} \geq 0$ is the maximum total improvement that can be obtained by splitting a failure tree rooted at $q' \succ q$ at any set of states $\preceq q$; $\check{\Delta}_{q|q'}$ is the maximum if this set includes $q$, and $\hat{\Delta}_{q|q'}$ is the maximum if this set does not include $q$.\footnote{A similar split into two cases---include $q$ or exclude $q$---is used in the well-known linear-time dynamic programming algorithm for finding the max-weighted independent set of vertices in a tree. In effect, these algorithms label each tree node with a bit saying whether or not to include it, subject to some constraints.  They resemble methods for refining the nonterminal labels of a parse tree (Petrov and Klein, 2007).}
The optimal split of $\tree$ then has total improvement $\sum_p \bar{\Delta}_{p\mid q_1}$ where $q_1$ is the root of $\tree$ and $p$ ranges over its parents.  

Tracing back through the derivation of this optimal improvement, one may determine which states were split to obtain it.  This is similar to following backpointers in the Viterbi algorithm.  Concretely, define
\begin{align}
\bar{S}_{q|q'} &=
\begin{cases}
\hat{S}_{q|q'} & \text{if }\bar{\Delta}_{q|q'}=\hat{\Delta}_{q|q'} \\
\check{S}_{q|q'} & \text{otherwise}
\end{cases} \label{eq:Sbar} \\
\check{S}_{q|q'} &=  \left(\textstyle\bigcup_p \bar{S}_{p|q}\right) \cup \{q\} \label{eq:Scheck} \\
\hat{S}_{q|q'} &= \left(\textstyle\bigcup_p \bar{S}_{p|q'}\right) \cup \emptyset \label{eq:Shat}
\end{align}
For example, $\bar{S}_{q|q'} \geq 0$ is the optimal set of states $\preceq q$ to split in a failure tree rooted at $q' \succ q$.
The optimal set of split points in $\tree$, not counting the original root $q_1$, is $S = \bigcup_p \bar{S}_{p|q_1}$ where, again, $p$ ranges over the parents of $q_1$.  Any of the unions written here can be enumerated by (recursively) enumerating the disjoint sets that are unioned together, without any copying to materialize the sets.

Concretely, we first work from the leaves down to the root: at each $q$, we compute and memoize all of the $\Delta$ quantities (quantities \labelcref{eq:deltabar}--\labelcref{eq:deltahat} for all $q' \succ q$), after first having done so at the parents of $q$. 
We then enumerate $S$ using the definitions \labelcref{eq:Sbar}--\labelcref{eq:Shat}, which recurse from the root back up to the leaves.  Thanks to the choice at \cref{eq:Sbar} based on the $\Delta$ quantities, this recursion enumerates only sets that are actually subsets of $S$.  In particular, it enumerates $\bar{S}_{q|q'}$ for just those $q' \succ q$ pairs such that $q'$ is the highest proper descendant of $q$ in the optimal set $\{q_1\} \cup S$.


The total runtime is dominated by \labelcref{eq:deltabar}--\labelcref{eq:deltahat} and is proportional to the number of $q' \succ q$ pairs in $\tree$.  Summed over all trees $\tree$, this is just the total height of all states in all failure trees, or equivalently $\sum_{q' \in \states} (\ancs(q')-1)$.  This resembles the failure term in the worst-case runtime of \cref{alg:new-backward-general}, but is much faster since it eliminates all factors that depend on $|\Sigma|$.  Thus, when a compatible order is not known (\cref{sec:toposort-heuristics}), taking the time to optimally split the failure trees may be worth the investment.  

\paragraph{Runtime analysis after static splitting}
To get a sense of how this improves the worst-case runtime, consider an idealized WFSA-$\phi$ where every state $q$ has  the same number of out-symbols, $\outsymbols(q)=s|\alphabet|$.  Furthermore, relax the runtime bound by replacing $\ancs(q)$ in the definition of $f(\tree)$ by the larger value $|\tree|$, so $f(\tree) \defeq s|\alphabet||\tree|^2$.  

This means when we split $\tree$ into $\tree_1,\ldots,\tree_K$, our earlier runtime expression $(K-1)|\alphabet| + 
\sum_{k=1}^K f(\tree_k) \log |\alphabet|$ becomes $(K-1)|\alphabet| + 
\sum_{k=1}^K s|\alphabet| |\tree_k|^2 \log |\alphabet|$, or more simply, $|\alphabet|(K-1 + (s \log |\alphabet|)\sum_{k=1}^K |\tree_k|^2)$.  For a given $K$, this is minimized when all $K$ trees have equal size $\frac{|\tree|}{K}$, yielding a minimum of 
\begin{equation}
|\alphabet|\,\left(K-1 + \frac{s \log |\alphabet|}{K}|\tree|^2\right)\label{eq:Kruntime}
\end{equation}
Setting the derivative with respect to $K$ to zero, we find that the optimal $K = |\tree| \sqrt{s \log |\alphabet|}$.

However, for a WFSA with sufficiently dense out-symbols, namely one with $s > \frac{1}{\log |\alphabet|}$, this asks to take $K > |\tree|$, which is impossible.  There the method will have to settle for $K = |\tree|$, splitting each state into its own failure tree.  This makes \cref{alg:new-backward-general} reduce to  \cref{alg:aggregation-backward}.

Conversely, for a WFSA with sufficiently sparse out-symbols, namely one with $s < \frac{1}{\maxfailuretreesize^2 \log |\alphabet|}$, the above formula asks to take $K < 1$ for all failure trees.  That is also impossible: the method will have to settle for $K=1$, not splitting $\tree$ at all.  This is the original version of \cref{alg:new-backward-general}.

In between these two extremes, we can take $K \approx |\tree| \sqrt{s \log |\alphabet|}$ as proposed above.  This makes the bound \labelcref{eq:Kruntime} on the contribution of failure tree $\tree$ to the runtime become $\bigO(|\alphabet| |\tree| \sqrt{s \log |\alphabet|})$.  Note that the $\sqrt{\ }$ term is $< 1$ because we are not too dense, so this may beat \cref{alg:aggregation-backward}.\jason{why doesn't it definitely beat it?  Is that just because we're using a looser bound in this section?}  It also beats the original \cref{alg:new-backward-general}: if we did not split the tree but kept $K=1$, the expression would give $\bigO(|\Sigma||\tree|^2 s \log |\alphabet|)$.  In short, splitting the tree avoids the quadratic worst-case cost of \cref{alg:new-backward-general}.  To put it another way, by eliminating the worst-case interaction among the $K$ trees, we have reduced from $\bigO(|\Sigma|K^2)$ to $\bigO(|\Sigma|K)$.  Recall that $K \geq 1$ since we are not too sparse, so this is again an improvement.

\section{Example of a Non-Suitable WFSA-$\phi$}

\begin{figure}[H]
    \centering
        \begin{tikzpicture}[node distance = 12 mm]
        \footnotesize
        \node[state, initial] (q1) [] { 1 }; 
        \node[state] (q2) [right =20mm of q1] { 2 }; 
        \node[state, accepting] (q3) [below right = of q1] { 3 }; 
        \draw[-{Latex[length=3mm]}] 
        (q1) edge[above, bend left=20] node{ $a$ } (q2) 
        (q2) edge[above, bend left=20, phitransitioncolor] node{ $\phi$ } (q1) 
        (q1) edge[below left, swap] node{ $b$ } (q3) 
        (q2) edge[below right, swap] node{ $a$ } (q3); 
        \end{tikzpicture}
        \caption{Example of a WFSA-$\phi$ where $\transitions \cup \transitions^\phi$ is not acyclic, yet its failure expanded transition set $\failureexpansion$ is.}
        \label{fig:failure-exp-acyclic-union-cyclic}
\end{figure}

\end{document}